\documentclass[fleqn,usenatbib]{mnras}

\usepackage{newtxtext,newtxmath}

\usepackage[T1]{fontenc}

\DeclareRobustCommand{\VAN}[3]{#2}
\let\VANthebibliography\thebibliography
\def\thebibliography{\DeclareRobustCommand{\VAN}[3]{##3}\VANthebibliography}

\usepackage{graphicx}	
\usepackage{amsmath}	
\usepackage[table]{xcolor}
\usepackage{array}

\newcommand{\dotdeg}{\rlap{.}^\circ}
\setcitestyle{notesep={ }} 
\newtheorem{proposition}{Proposition}

\title[Cosmic Multipoles I]{%
    Cosmic Multipoles in Galaxy Surveys Part I:\\
    How Inferences Depend on Source Counts and Masks}

\author[Oayda et al.]{
Oliver T. Oayda,$^{1}$\thanks{E-mail: oliver.oayda@sydney.edu.au}
Vasudev Mittal,$^{1}$
Geraint F. Lewis$^{1}$
\\
$^{1}$Sydney Institute for Astronomy, School of Physics A28, The University of Sydney, NSW 2006, Australia
}

\date{Accepted 2024 December 15. Received 2024 December 12; in original form 2024 November 3}

\pubyear{2024}

\begin{document}
\label{firstpage}
\pagerange{\pageref{firstpage}--\pageref{lastpage}}
\maketitle

\begin{abstract}
We present a new approach to constructing and fitting dipoles
and higher-order multipoles in
synthetic galaxy samples over the sky.
Within our Bayesian paradigm,
we illustrate that this technique is robust to masked skies,
allowing us to make credible inferences about
the relative contributions of each multipole.
We also show that dipoles can be recovered
in surveys with small footprints,
determining the requisite source counts required
for concrete estimation of the dipole parameters.
This work is motivated by recent probes
of the cosmic dipole in galaxy catalogues.
Namely, the kinematic dipole of the Cosmic Microwave Background,
as arising from the motion of our heliocentric frame
at $\approx 370\ \text{km}\,\text{s}^{-1}$,
implies that an analogous dipole should be observed
in the number counts of galaxies in flux-density-limited samples.
Recent studies have reported a dipole aligning with
the kinematic dipole but with an anomalously large amplitude.
Accordingly, our new technique will be important
as forthcoming galaxy surveys are made available
and for revisiting previous data.
\end{abstract}




\section{Introduction}
In the current cosmological paradigm,
the temperature anisotropies in the Cosmic Microwave Background (CMB)
-- as arising from matter-photon interactions in the early universe --
are thought to be the originators of structure in the late universe
\citep{coles2002}. However,
these small-scale fluctuations are two orders of magnitude smaller
than the temperature dipole measured in the CMB.
The standard interpretation is that this dipole
arises from the motion of our heliocentric reference frame
through the Universe with respect to the `cosmic rest frame',
or the rest frame of the CMB \citep{peebles1968};
thus, this dipole is termed the `kinematic dipole'.
Under the assumption of the cosmological principle,
Lorentz boosting to the frame in which the CMB exhibits no dipole
should correspond to positioning oneself in the frame where,
on very large scales,
the Universe is homogeneous and isotropic \citep{harrison2000}.

Whether or not this is so can be tested.
The motion of our heliocentric frame,
as ascertained from the kinematic dipole,
should impact other observables.
This includes the distribution of matter in large-scale
surveys of radio galaxies and quasars,
in which the source density per patch of sky
is expected to be the sum of a monopole and dipole signal
\citep{ellis1984}.
The expected magnitude of the latter signal is determined by the kinematic dipole
and the underlying properties of the source population.
However,
recent studies (see Section~\ref{sec:background} and references therein)
have found evidence for a dipole in source density aligning with the kinematic dipole
but larger in amplitude.
Across these studies,
typically the amplitude is between 2 to 3 times the size expected
from the kinematic dipole.
We refer to this growing anomaly as the `dipole tension'.

Such an anomaly raises several questions:
do we interpret the excessive dipole as a breakdown of
the assumption of homogeneity and isotropy,
and therefore a breakdown of the cosmological principle;
or, is there a systematic effect which has as of yet
not been adequately accounted for?
One critical issue is the statistical framework
and methodology used to probe the source density variation
in a source catalogue.
Certain approaches include decomposing the source density map --
a representation of the number of sources per cell over the celestial sphere --
into spherical harmonics,
evaluating the harmonic coefficients for the $\ell=1$ mode (dipole)
and potentially higher order multipoles
\cite[see e.g.][]{abghari2024}.
This has led to the claim that leakage from higher order multipoles
impacts the measurement of the dipole amplitude for partial sky maps,
which at the least throws some doubt on the scale
of the amplitude uncertainties.
Regardless of the approach used,
understanding the higher order effects present in the data is important
for accurate and credible inferences about the scale and direction of
the dipole in source density.

With this context in mind,
in this work we generate synthetic source catalogues onto which dipole,
quadrupole and octupole signals are imprinted.
Instead of relying on spherical harmonic functions, we simulate and fit
these functions by specifying only an amplitude and directional unit vectors.
This technique for constructing multipoles has had, to our knowledge,
limited application in the literature -- and has not been used in
the context of the kinematic dipole studies.
We then deploy a Bayesian statistical approach,
determining how the presence or absence of either multipole
impacts the conclusions inferred.
We also determine how these conclusions change with different masks,
as well as with synthetic samples at varying source densities.
The net effect of this exploration is a deeper understanding
of the properties and sensitivities of our statistical approach.
We illustrate how our methodology is robust at disentangling
the individual contributions from a dipole, quadrupole and octupole
without cross-talk between them.
We also showcase how we can make credible inferences on incomplete skies where
more than half of the sky is masked.
In doing so, we demonstrate the advantages of our approach over other techniques,
like the use of estimators under a frequentist paradigm.
In fact, certain estimators prevalent in the literature
suffer from an intrinsic bias in both dipole direction
and magnitude.
This bias is a result of imperfections in the data
such as shot noise and insufficient sky coverage.
In a forthcoming companion study,
we will explore popular estimators to assess the presence of intrinsic
biases due to these survey properties \citep{mittalPREP}.

Our paper is set out as follows:
in Section~\ref{sec:background},
we give a brief summary of the literature surrounding this study;
in Section~\ref{sec:approach},
we describe our approach,
including the method of generating synthetic samples
and our statistical regime;
in Section~\ref{sec:results}, we present our findings,
and; in Section~\ref{sec:discussion_conclusions},
we discuss the significance of these results
with an eye to future measurements and studies.

\section{Background}
\label{sec:background}
Measurements of the kinematic dipole have characterised our heliocentric motion
with a speed of $v_{\text{CMB}} = 369.82\pm0.11\,\text{km}\,\text{s}^{-1}$
and in the direction $(l,b) = (264\dotdeg021, 48\dotdeg253)$
in Galactic coordinates \citep{planck2020}.
If this dipole is wholly kinematic in origin,
then we can compute the expected effect it will have on galaxy samples
and compare it to what we observe empirically.

Such a test was formulated by \citet{ellis1984}
using only special relativistic arguments,
coupled with a handful of assumptions
about the underlying source population.
Namely,
suppose that the spectral energy distribution of the sources
follows a power law described by the spectral index $\alpha$,
where $S_\nu \propto \nu^{-\alpha}$.
In addition,
assume that the total number of sources
above some limiting flux density follows a power law
characterised by the exponent $x$; that is,
$N (> S_\nu) \propto S_\nu^{-x}$.
In the rest frame of the sources,
an observer perceives an isotropic and homogeneous
distribution of objects.
Transforming to a frame moving with respect to the source background,
which we denote with primed variables,
the flux density of a source in some passband is Lorentz boosted.
This means that $S_\nu' = S_\nu\delta^{1+\alpha}$,
where $\delta = \gamma(1 + \beta \cos \theta)$
for Lorentz factor $\gamma$, $\beta = v / c$
and angle $\theta$ between
the source and the observer's direction of motion.
In addition,
the element of solid angle transforms as
$d\Omega' = d \Omega \delta^{-2}$,
which describes relativistic aberration.
Taking into account the cumulative flux density distribution
and given $v \ll c$,
\citet{ellis1984} determined that -- at first order -- we should perceive
a dipolar variation in source density with amplitude
\begin{equation}
    \mathcal{D} = (2 + x(1 + \alpha)) \beta.
\end{equation}
For example,
using $v_\text{CMB} \approx 370\,\text{km}\,\text{s}^{-1}$
and taking typical values of $x = 1$ and $\alpha = 0.75$,
the expected dipole amplitude is $\mathcal{D} \approx 4.6 \times 10^{-3}$.
This means that we would expect a
$\approx 0.5\%$ increase in source density
from the mean directly ahead of our motion
(the pole of the forward hemisphere),
and a $\approx 0.5\%$ decrease in source density
directly behind our motion.
Though this effect is subtle, with sufficient sample sizes it is
in principle detectable.

The \citet{ellis1984} test has been carried out
with catalogues of radio galaxies,
as well as catalogues of quasars recorded in the near-IR
and optical regimes.
As a high-level overview, the current literature prefers a
source density dipole that points in roughly
the same direction as the kinematic dipole,
but has an amplitude in excess
\citep[see e.g.][but see also \citealt{blake2002, mittal2024, wagenveld2024}]{
    gibelyou2012, rubart2013, colin2017, bengaly2018,
    siewert2021, secrest2021, secrest2022, singal2023,
    wagenveld2023, oayda2024}.
This has led some to view the dipole tension
in the context of other anomalies in $\Lambda$CDM,
with an eye to new physics that
builds upon our current paradigm \citep[see e.g.][]{Peebles_2022, aluri2023}.
It also lends itself to the importance of trying other
tests of the cosmological principle,
such as those using
Type Ia SNe \citep[see e.g.][]{hu2024}
or kinematically-induced time dilation \citep{oayda2023}.

However,
one must also ask if there are systematic effects
which contribute -- possibly in whole or in part --
to the dipole tension.
For example,
some authors have proposed that source evolution
is an important factor that has hitherto been neglected
\citep[see][]{dalang2022, gaundalin2023}.
Namely, the $\alpha$ and $x$ we determine
are in effect averaged quantities over the sample redshift distribution,
and care must be taken in accounting for how they change with redshift
\citep[but see][which claims this effect
        does not impact the dipole measure]{vonHausegger2024}.
Another issue is local clustering, in which
the inhomogeneous distribution of matter in our cosmic neighbourhood
is expected to have a dipole component.
This is undesirable insofar that the dipole we wish to measure
is special relativistic in origin,
consequent on the motion of our heliocentric frame,
and not the dipole in nearby cosmic structure.
Local clustering was conceived as a possible contaminant
for the NRAO VLA Sky Survey \citep[NVSS;][]{nvss-survey}
in \citet{blake2002}, an early study of that sample. 
In \citet{oayda2024},
local clustering was shown to impact the dipole measurement
in NVSS and the Rapid ASKAP Continuum Survey \citep[RACS;][]{racs-original}
by as much as 10\%--15\%.
Nonetheless,
this is still insufficient to fully explain
the scale of the dipole tension.

One peripheral question relates to the methods
used to infer the source count dipole.
For example,
\citet{blake2002} relied on the decomposition
of the number count density function over the celestial sphere
into spherical harmonics $Y_{\ell m}$ (up to $\ell=3)$,
as well as the earlier work of \citet{baleisis1998}.
So, for some density field $\sigma(\phi,\theta)$
defined over the celestial sphere by azimuthal and polar angles
$\phi$ and $\theta$ respectively, we have that
\begin{equation}
    \sigma(\phi, \theta) = \sum_{\ell} \sum_{m} a_{\ell m} Y_{\ell m} (\phi, \theta)
\end{equation}
where $a_{\ell m}$ are the spherical harmonic coefficients.
\citet{blake2002} then used a frequentist approach,
comparing a dipole model to the observed coefficients
with a chi-squared ($\chi^2$) test.

Incomplete sky coverage causes issues for
spherical harmonic decompositions.
Usually, portions of the sky are masked due to survey coverage limits
and/or contamination from the Galactic plane.
This is an issue since the harmonic coefficients
are found by integrating over a complete sky.
More specifically,
finding the coefficients relies on the fact that
the harmonic functions are orthonormal,
which is broken when the integral is no longer bounded over the entire sphere
\citep[see e.g.][for a recent discussion on this issue]{abghari2024}.
Consequently, performing a decomposition on an incomplete sky
introduces coupling effects between different harmonic modes,
biasing the inferred angular power spectrum.
This bias needs to be accounted for if one wants to recover
a genuine estimate of the angular power spectrum of a galaxy survey,
and hence the power of the dipole ($\ell=1$) mode.

\citet{abghari2024} claimed that mode mixing is a genuine
concern for the analysis of CatWISE2020 \citep{marocco2021}
in \citet{secrest2021}.
In the latter study, the authors reported a $5\sigma$ tension between the
expected and inferred dipole amplitudes.
The claim of \citet{abghari2024} is contingent on the statement
that the estimator used in \citet{secrest2021},
a least-squares estimator minimising the error between
the observed cell counts and the model counts
from monopole and orthogonal dipole templates,
suffers from mode coupling.
This is because the estimator sums $\ell=1$ harmonic templates,
and the mask deployed in \citet{secrest2021} covers about $50\%$ of the celestial sphere.
The CatWISE2020 sample also suffers from a known ecliptic bias,
with elevated source density around the ecliptic equator
and diminished densities at the poles.
This forms a strong quadrupole ($\ell=2$) mode.
But, as \citet{abghari2024} contends, even after correcting for this effect
there are still non-negligible higher order harmonics.
The effect this has on inference of the dipole ($\ell=1$) amplitude is not
immediately obvious, but it could suggest the scale of the uncertainties
has been underestimated.

We will keep this issue as a running theme as we present our methodology
and results.
Specifically, we will turn our attention to the case of partial (masked)
skies which exhibit higher order multipoles, in addition to a dipole.
As mentioned earlier,
we do not rely on a spherical harmonic decomposition of the
source count density map.
Instead, we create specific parametric models describing a dipole, quadrupole and
octuople and fit them to synthetic samples using Bayesian statistics.
In fact, our approach is generalisable to higher order multipoles -- the key
issue being the additional computational overhead.
We show that this approach avoids any issues of power leakage between harmonic modes.

\section{Approach}
\label{sec:approach}
\subsection{Mathematical Underpinning}
\subsubsection{Monopole}
In our study,
a monopole simply describes an average source density.
By definition,
the monopole has no directional dependence but a constant scalar
value for all patches of sky.
Thus, we define the monopole signal as
\begin{equation}
    f_{\text{mono.}} = \bar{\mathcal{N}} \label{eq:monopole_expect}
\end{equation}
for average source density $\bar{\mathcal{N}}$ (in units of sources per cell),
and the expected count $\mathcal{N}_i$ in cell $i$ is
\begin{equation}
    \mathbb{E}[\mathcal{N}_i] = \bar{\mathcal{N}}.
\end{equation}

\subsubsection{Dipole}
To represent a dipole in source density,
we construct the vector $\mathbf{d}$ with magnitude $\mathcal{D}$
and pointing in some direction $(l^\circ, b^\circ)$ in Galactic coordinates.
Then, we define the `pixel vector' as the unit vector $\hat{\mathbf{p}}$
pointing towards an arbitrary patch of sky.
The dipole signal is the dot product of the dipole vector
and the pixel vector, where
\begin{equation}
    f_{\text{dip.}} = \mathbf{d} \cdot \hat{\mathbf{p}}
                    = d_j \hat{p}_j
                    = \mathcal{D} \cos \theta.
\end{equation}
$\theta$ is the angle between the dipole vector and the patch of sky,
and after the second equality, we have used index notation
(with summation assumed over the repeated indices $j$).
We introduce this notation now with the intention of generalising it
for higher order multipoles later in this section.
As an example,
if $\mathcal{D} = 0.1$,
then at the pole of the forward hemisphere where $\theta = 0$
we have that $f_\text{dip.} = 0.1$,
i.e. the dipole contributes a 10\% enhancement
in source density in that direction.

To compute the expected count in any cell,
we need information about the underlying source density (the monopole)
as well as the dipole.
Thus, if we suppose the highest order multipole in a sample is a dipole,
we have
\begin{equation}
    \begin{aligned}
        \mathbb{E}[\mathcal{N}_i] = f_\text{mono.} + f_\text{mono.}f_\text{dip.}
                          &= \bar{\mathcal{N}} + \bar{\mathcal{N}}(
                            \mathcal{D} \cos \theta_i)\\
                          &= \bar{\mathcal{N}} (1 + \mathcal{D} \cos        \theta_i) \label{eq:dipole_expect}
    \end{aligned}
\end{equation}
where $\theta_i$ is the angle between the vector pointing to cell $i$
and the dipole vector.

\subsubsection{Quadrupole}
We cannot represent a quadrupole with a single vector,
as was the case for the dipole.
Instead,
one must construct the quadrupole tensor $Q$,
which in this case is a traceless symmetric $3\times3$ matrix.
We refer to the approach expounded in the foregoing sections as the
`traceless symmetric tensor approach', which has had key applications
in gravitational waves, studies of low-$\ell$ CMB harmonics,
galaxy spin isotropy measures
and electromagnetism
\citep[see e.g.][]{pirani1965, schwarz2004, land2008, guth2012_l8, guth2012_l9}.%
\footnote{%
    For the CMB studies, \citet{schwarz2004} references \citet{copi2024}
    as the source of the unit vector or traceless symmetric tensor approach,
    but \citet{weeks2004} points out that the method in fact originates
    with Maxwell's \textit{Treatise on Electricity and Magnetism} (1873).
}
To construct the quadrupole tensor,
we first take the outer product of two unit vectors,
$Q' = \hat{\mathbf{a}} \otimes \hat{\mathbf{b}}$.
If the components of each vector are
$\mathbf{\hat{a}} = (x_1, y_1, z_1)$ and
$\mathbf{\hat{b}} = (x_2, y_2, z_2)$,
then the explicit matrix representation of $Q'$ is
\begin{equation}
    Q' = 
    \begin{pmatrix}
        x_1 x_2 & x_1 y_2 & x_1 z_2 \\
        y_1 x_2 & y_1 y_2 & y_1 z_2 \\
        z_1 x_2 & z_1 y_2 & z_1 z_2 
    \end{pmatrix},
\end{equation}
or, in index notation, an element of $Q'$ is $Q'_{jk} = \hat{a}_j \hat{b}_k$.
We will now use index notation for subsequent expressions.
A symmetric matrix $Q^*$ can be constructed from $Q'$ by averaging over
all permutations of its indices.
With only 2 indices, this amounts to
\begin{equation}
    Q^*_{jk} = \frac{1}{2} \left( Q'_{jk} + Q'_{kj} \right). \label{eq:quad-sym}
\end{equation}
We then compute the trace of this matrix ($Q_{ll}$)
and render it traceless through
\begin{equation}
    \hat{Q}_{jk} = Q^*_{jk} - \frac{Q^*_{ll}}{3}
\end{equation}
Lastly, we introduce the scalar term $\mathcal{Q}$ which magnifies or diminishes
the elements of $\hat{Q}$,
identifying this as the quadrupole amplitude.\footnote{Whereas
$\mathcal{D}$ encodes the difference from the mean to maximal
values of the dipole map (e.g. $\mathcal{D} = 0.007$ means a
0.7\% enhancement in source density or signal in the forward hemisphere), $\mathcal{Q}$ encodes the difference between the minimal and maximal values
of the quadrupole signal. To see this, consider the top row of Fig.~\ref{fig:sample-templates}, noting the values in the colour bar there.}
Thus,
\begin{equation}
    Q_{jk} = \mathcal{Q} \hat{Q}_{jk}.
\end{equation}
$Q$ has five independent components, since it is by construction symmetric
and traceless.
Analogously, we need only five parameters to define a quadrupole:
the position of each vector on the unit sphere (four parameters),
and the quadrupole amplitude (one parameter).
The quadrupole signal is then the matrix-vector product
\begin{equation}
      f_\text{quad.} = Q_{jk} \hat{p}_j \hat{p}_k
                     = \mathcal{Q} \left( \hat{Q}_{jk} \hat{p}_j \hat{p}_k
                        \right).
\end{equation}
This yields a scalar for each position on the sky $\mathbf{\hat{p}}$.
Then,
the expected count in a cell
(assuming only a quadrupole is imprinted on the sample) is
\begin{equation}
    \mathbb{E}[\mathcal{N}_i] = f_\text{mono.} + f_\text{mono.}f_\text{quad.}
                              = \bar{\mathcal{N}}
                              \left(
                                    1 + \mathcal{Q}
                                    \left(
                                        \hat{Q}_{jk} \hat{p}_j \hat{p}_k
                                    \right)
                              \right).
                                \label{eq:quadrupole_expect}
\end{equation}

\subsubsection{Octupole}
We can begin to generalise the foregoing process for higher order multipoles.
With the dipole and quadrupole being defined by one and two unit vectors
(plus an amplitude) respectively, we can construct a $3\times3\times3$ octupole
tensor $O$ with three unit vectors.
In particular, for three unit vectors $\hat{\mathbf{a}}$, $\hat{\mathbf{b}}$
and $\hat{\mathbf{c}}$,
the octupole tensor in non-symmetric and non-zero trace form is
\begin{equation}
    O'_{jkl} = \hat{a}_j \hat{b}_k \hat{c}_l.
\end{equation}
As hinted at before, symmetry implies that the elements of a tensor
are the same under any permutation of its indices; thus,
we again average over all permuted tensors (cf. \eqref{eq:quad-sym}).
Explicitly, denote the set of all possible permutations of the indices $jkl$
by $S$, which has 6 elements indexed by $m$. Then
\begin{equation}
    O^{*}_{jkl} = \frac{1}{3!} \sum_{m=1}^6 O'_{S_m} = O'_{(jkl)}.
\end{equation}
Note that the parentheses in the subscript after the last equality indicates
symmetrisation over the enclosed indices; we reuse this notation at a later point.
Trace, which we denote with $T$, generalises to a contraction of a tensor's indices.
Thus,
\begin{equation}
    T_l = \delta_{jk} O^*_{jkl} = O^*_{jjl},
\end{equation}
where $\delta_{jk}$ is the Kronecker delta.
Note that the trace is now a vector with three components and not a scalar.
Since the tensor $O^*$ is symmetric, $T_l$ is the same for a permutation
of the indices $jjl$.
To render $O^*$ traceless, we compute
\begin{align}
    \hat{O}_{jkl} &= O^*_{jkl} -
        \frac{1}{5} \left( \delta_{jk}O^*_{mml}
                            + \delta_{jl}O^*_{mmk}
                            + \delta_{kl}O^*_{mmj} \right) \\
                            \label{eq:oct-zero-trace}
        &= O^*_{jkl} - \frac{3}{5} \delta_{(jk} O^*_{l)mm}.
\end{align}
Note that the brackets overflow from the Kronecker delta 
$\delta$ to $O^*$; i.e., the indices $jkl$ are symmetrised over.
One can verify that this yields the desired result of zero trace
by contracting both sides of \eqref{eq:oct-zero-trace} with $\delta_{jk}$.
Lastly, we introduce $\mathscr{O}$ as the octupole amplitude, defined by
\begin{equation}
    O_{jkl} = \mathscr{O} \hat{O}_{jkl}.
\end{equation}
By construction, the octupole tensor $O$ has seven independent components ---
initially starting with 27 components, imposing symmetry reduces the count to 10,
and setting the trace to be zero reduces it again to 7.
The octupole signal is given by
\begin{equation}
    f_{\text{oct.}} = O_{jkl} \hat{p}_j \hat{p}_k \hat{p}_l
            = \mathscr{O} \left( \hat{O}_{jkl} \hat{p}_j \hat{p}_k \hat{p}_l \right)
\end{equation}
and the expected count in cell $i$ by
\begin{equation}
    \mathbb{E}[\mathcal{N}_i] = f_{\text{mono.}} + f_{\text{mono.}} f_{\text{oct.}}
    = \bar{\mathcal{N}} \left( 1 + \mathscr{O} \left(
                    \hat{O}_{jkl} \hat{p}_j \hat{p}_k \hat{p}_l \right) \right).
\end{equation}

\subsubsection{General Multipole}
For the sake of completeness, we also give the expression for a general
multipole of any order $\ell$.
This involves first symmetrising over the rank-$\ell$ tensor constructed from
the $\ell$ unit vectors, and then imposing tracelessness. Explicitly, we have
\begin{equation}
    M'_{i_1 i_2 \ldots i_\ell}
        = (\hat{n}_1)_{i_1} (\hat{n}_2)_{i_2} \ldots (\hat{n}_\ell)_{i_\ell}
        = \prod_{k=1}^{\ell} (\hat{n}_k)_{i_k}
\end{equation}
where $i_1$ denotes `index 1' and $\mathbf{\hat{n}}_1$ denotes the first unit vector
of all $\ell$ vectors. Again using the notation that the symmetrised tensor
$M^*_{i_1 i_2 \ldots i_\ell} = M'_{(i_1 i_2 \ldots i_\ell)}$, we have that the traceless
symmetric tensor for the multipole is
\begin{equation}
    \hat{M}_{i_1 i_2 \ldots i_\ell} = \sum_{k=0}^{\lfloor \ell / 2 \rfloor}
                            (-1)^{k} \frac{\binom{\ell}{k} \binom{\ell}{2k}}{\binom{2\ell}{2k}}
                            \delta_{(i_1 i_2} \ldots \delta_{i_{2k-1} i_{2k}}
                            M^*_{i_{2k+1} \ldots i_{l}) j_1 j_1 \ldots j_k j_k}.
        \label{eq:general_traceless_tensor}
\end{equation}
This expression can be found in \citet{pirani1965},
and of course reduces to the expressions given above for the dipole ($\ell=1$),
quadrupole ($\ell=2$) and octupole ($\ell=3$) tensors.
For an intuition as to why this expression is correct,
it is instructive to write out the explicit cases for multipoles up to $\ell=4$,
as is done in \citet{guth2012_l9}.
Like the earlier cases, we introduce the multipole amplitude $\mathscr{M}$ by
\begin{equation}
    M_{i_1 i_2 \ldots i_\ell} = \mathscr{M} \hat{M}_{i_1 i_2 \ldots i_\ell}.
\end{equation}
Lastly, the multipole signal and expected cell count are given by
\begin{equation}
    f_{\text{mult.}} = M_{i_1 i_2 \ldots i_l}
                        \hat{p}_{i_1} \hat{p}_{i_2} \ldots \hat{p}_{i_\ell}
                     = \mathscr{M} \left(
                        \hat{M}_{i_1 i_2 \ldots i_\ell} \hat{p}_{i_1} \hat{p}_{i_2} \ldots \hat{p}_{i_\ell}
                            \right) \label{eq:multi_signal}
\end{equation}
and
\begin{equation}
    \mathbb{E} [\mathcal{N}_i] = f_{\text{mono.}} + f_{\text{mono}.} f_{\text{mult.}}
                               = \bar{\mathcal{N}} \left( 1 + \mathscr{M}
                                    \left(
                                        \hat{M}_{i_1 i_2 \ldots i_\ell} \hat{p}_{i_1} \hat{p}_{i_2} \ldots \hat{p}_{i_\ell}
                                    \right)
                                \right)
\end{equation}
respectively.
The main challenge involved with this approach is the computational complexity.
To see this, note that symmetrisation is being performed over all
permutations of a tensor's indices -- even at the stage of
imposing zero trace.
With an $\ell=10$ multipole, this amounts to $10! \approx 3.6$ million
permutations.

We give an example of the multipole signal projected onto the sky
in Fig.~\ref{fig:n_multipoles}. Here, we chose to sum together
the contributions from multipoles of order $\ell=1$, $\ell=2$, $\ell=3$,
$\ell=6$, $\ell=7$ and $\ell=8$ using \eqref{eq:multi_signal}.
After randomly generating each unit vector, we used the following multipole amplitudes:
$\mathscr{M}_1 = 0.007$, $\mathscr{M}_2 = 0.014$, $\mathscr{M}_3 = 0.03$,
$\mathscr{M}_6 = 0.5$, $\mathscr{M}_7 = 1$ and $\mathscr{M}_8 = 2$.
We also performed a harmonic analysis
on this map with \textsc{healpy's} \verb|anafast| function,
plotting the angular power spectrum for $1 \leq \ell \leq 10$ in the same figure.
There, the coefficients $C_4$ and $C_5$ are zero, as expected.

\subsection{Simulations}
\subsubsection{Catalogue Templates}
To generate our synthetic catalogues,
we first divide the celestial sphere into equal-area pixels
using the \textsc{healpix}\footnote{\url{https://healpix.sourceforge.io/}}
procedure implemented in the \textsc{python} package \textsc{healpy} \citep{Gorski2005, Zonca2019}.
If we have a sample of sources (points) distributed over
the celestial sphere,
the process of binning them into pixels
of fixed area describes a Poisson point process --
assuming the location of each source is independent.%
\footnote{%
    In practice, this assumption is broken where one source may
    be coupled with additional sources, for example in radio
    continuum surveys where multi-component sources are typical.
    This challenge is discussed in \citet{oayda2024}
    in the context of cross-matching radio sources to their optical
    counterpart (see section 6.3 therein), and is an ongoing issue.
}
Thus, the probability distribution for the number of points
in a given cell is parameterised only by
the rate parameter $\lambda$ of the Poisson distribution.

\begin{figure*}
    \centering
    \includegraphics[width=0.9\linewidth]{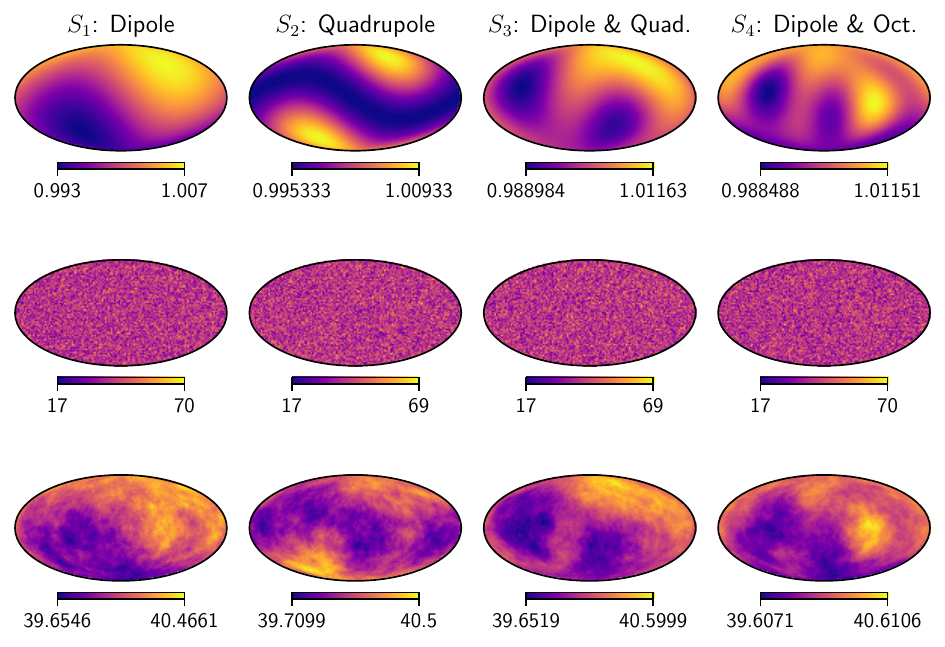}
    \caption{Visualisation of our catalogue templates projected onto
    the sky in Galactic coordinates (Mollweide).
    \textit{Top row:} The raw signals, as listed in Section~\ref{subs:param_optim}.
    \textit{Middle row:} One realisation of a possible density map
    $\bar{\mathcal{N}} = 40$ sampled from the above signal map
    (each cell is a random deviate drawn from a Poisson
    distribution specific to that cell).
    \textit{Bottom row:} The above density map smoothed with a 1 steradian
    moving average, illustrating the underlying large-scale features.
    \textit{Left column:} Dipole sample $S_1$ with amplitude $\mathcal{D} = 0.007$.
    \textit{Middle-left column:} Quadrupole sample $S_2$ with amplitude $\mathcal{Q} = 0.014$.
    \textit{Middle-right column:} Dipole and quadrupole sample $S_4$ with
        amplitude $\mathcal{D} = 0.007$ and $\mathcal{Q} = 0.014$.
    \textit{Right column:} Dipole and octupole sample $S_4$ with
        amplitude $\mathcal{D} = 0.007$ and $\mathscr{O} = 0.03$.}
    \label{fig:sample-templates}
\end{figure*}

The value of the rate parameter depends on what signal
we imprint onto the sample.
For example,
if the points are distributed according to a dipole,
then $\lambda_i = \bar{\mathcal{N}} (1 + \mathcal{D} \cos \theta_i)$
(see \eqref{eq:dipole_expect}).
This is the rate parameter for the $i$-th pixel on the sky.
Thus,
we may draw a number count from the Poisson distribution of each cell,
yielding a pixel density map.
That is,
the $i$-th cell count $\mathcal{N}_i$ is drawn from the probability distribution
\begin{equation}
    P( \mathcal{N}_i | \lambda_i )
        = \frac{\lambda_i^{\mathcal{N}_i} e^{-\lambda_i}}{\mathcal{N}_i!}.
\end{equation}
Shot noise is an inevitable (but desirable) consequence of this approach,
describing the statistical fluctuations in density from pixel to pixel
due to the binning process and finite source counts.
In this manner,
we generate four types of synthetic samples, as described below.
\begin{itemize}
    \item $S_1$: A dipole sample using \eqref{eq:dipole_expect}.
    \item $S_2$: A quadrupole sample using \eqref{eq:quadrupole_expect}.
    \item $S_3$: A sample consisting of a quadrupole and dipole signal.
    \item $S_4$: A sample consisting of an octupole and dipole signal.
\end{itemize}
For sample $S_3$,
we simply add the dipole and quadrupole terms together
such that the rate parameter for the $i$-th pixel is
\begin{equation}
    \begin{aligned}
        \lambda_i &= f_{\text{mono.}} + f_{\text{mono.}}f_{\text{dip}}
                    + f_{\text{mono.}}f_{\text{quad.}} \\
                  &= \bar{\mathcal{N}} \left(
                        1 + d_j \hat{p}_j
                    + \mathcal{Q} \left(
                        \hat{Q}_{jk} \hat{p}_j \hat{p}_k
                    \right) \right).
    \end{aligned}
\end{equation}
Similarly, for sample $S_4$, we have
\begin{equation}
    \begin{aligned}
        \lambda_i &= f_{\text{mono.}} + f_{\text{mono.}}f_{\text{dip}}
                    + f_{\text{mono.}}f_{\text{oct.}} \\
                  &= \bar{\mathcal{N}} \left(
                        1 + d_j \hat{p}_j
                    + \mathscr{O} \left(
                    \hat{O}_{jkl} \hat{p}_j \hat{p}_k \hat{p}_l \right) \right).
    \end{aligned}
\end{equation}
We show a visualisation of these templates in Fig.~\ref{fig:sample-templates}.
This figure also indicates the dipole and/or quadrupole amplitudes
chosen for each sample.
We list all the chosen parameter values in Table~\ref{tab:sample_params}.
\begin{table}
    \centering
    \setlength{\tabcolsep}{3pt}
    \rowcolors{2}{gray!25}{white}
    \renewcommand{\arraystretch}{1.2}
    \begin{tabular}{l >{\centering\arraybackslash}p{0.4\linewidth} p{0.4\linewidth}}
        \hline
         Sample & Amplitudes & Directions \\
         \hline
         $S_1$ & $\mathcal{D} = 0.007$
               & $(l,b) = (l_{\text{CMB}}, b_{\text{CMB}})$ \\
         $S_2$ & $\mathcal{Q} = 0.014$
               & $(l_1,b_1) = (96\dotdeg3, -60\dotdeg2)$\newline
                 $(l_2,b_2) = (96\dotdeg3, -60\dotdeg2)$ \\
         $S_3$ & \qquad$\mathcal{D} = 0.007$\newline $\mathcal{Q} = 0.014$
               & $(l,b) = (l_{\text{CMB}}, b_{\text{CMB}})$\newline
                 $(l_1,b_1) = (302\dotdeg9, -27\dotdeg1)$\newline
                 $(l_2,b_2) = (122\dotdeg9, 27\dotdeg1)$ \\
         $S_4$ & \qquad$\mathcal{D} = 0.007$\newline $\mathscr{O} = 0.03$
               & $(l,b) = (l_{\text{CMB}}, b_{\text{CMB}})$\newline
                 $(l_1,b_1) = (118\dotdeg0, -11\dotdeg6)$\newline
                 $(l_2,b_2) = (307\dotdeg8, -47\dotdeg4)$\newline
                 $(l_3,b_3) = (49\dotdeg4, 1\dotdeg6)$ \\\hline
    \end{tabular}
    \caption{Dipole, quadrupole and octupole parameters used for each sample.
    $(l,b)$ denotes the dipole direction, whereas ($l_1, b_1$), ($l_2, b_2$), etc.
    denote the higher order multipole unit vectors. Since the two
    quadrupole unit vectors in sample $S_2$ are identical,
    the sample is azimuthally symmetric about the axis defined by the vectors.}
    \label{tab:sample_params}
\end{table}
Note that the two unit vectors used to construct the quadrupole in sample
$S_2$ are the same.
In this symmetric traceless tensor formalism,
this choice amounts to enforcing symmetry about the identical axis represented
by each vector.

\subsubsection{Catalogue Permutations}
\label{subs:permutations}
\begin{figure}
    \centering
    \includegraphics[width=0.80\linewidth]{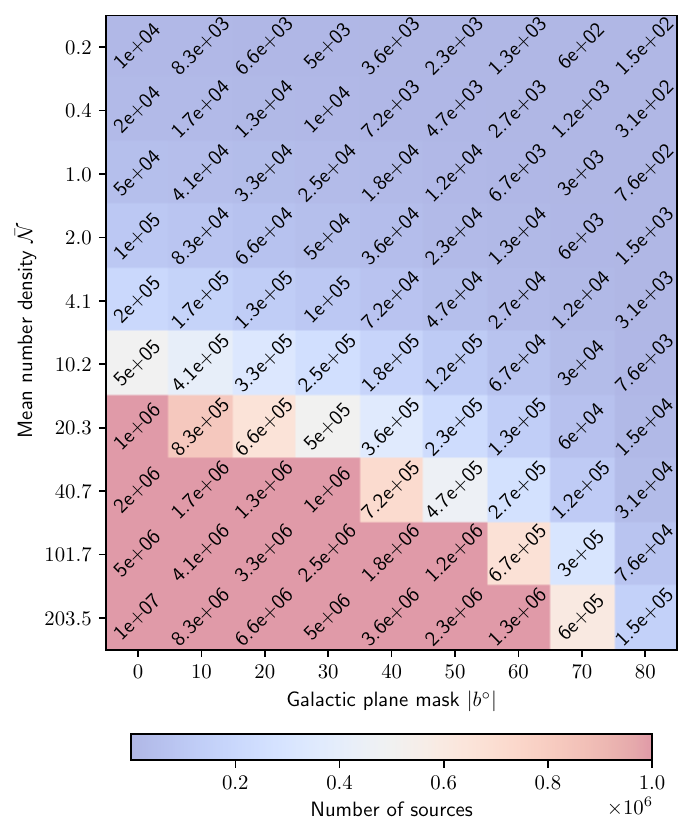}
    \caption{Expected number of sources in each of our synthetic
             catalogue permutations. The colour scale indicates
             the number of sources, with red being higher and blue being lower.
             The actual values are shown at the centre of each cell.
             \label{fig:expected_count_map}}
\end{figure}

Starting from the templates described above,
we generate sample variations by changing the total number
of sources $N$ and the choice of mask.
Namely, we generate samples with values of $N$ up to 10,000,000.
We also try masking the Galactic plane in increments of 10$^\circ$.
The specific values we nominated,
as well as the expected number of sources in each catalogue permutation,
are shown in Fig.~\ref{fig:expected_count_map}. 

\subsection{Statistical Regime}
\label{sub:stats-regime}
Our statistical procedure is the same as was used
in \citet{mittal2024} and \citet{oayda2024}.
Namely,
we rely on Bayesian inference to convert
prior assumptions or beliefs into statements conditioned on the data, where
\begin{equation}
    P( \Theta | \mathbf{D}, M )
        = \frac{\mathcal{L}(\mathbf{D} | \Theta, M) \pi(\Theta | M)}
        {\mathcal{Z}(\mathbf{D} | M)} \label{eq:bayes-theorem}
\end{equation}
for dataset $\mathbf{D}$, parameters $\Theta$ and model $M$,
as well as likelihood function, prior function and marginal likelihood
$\mathcal{L}$, $\pi$ and $\mathcal{Z}$ respectively.

\subsubsection{Parameter Optimisation}
\label{subs:param_optim}
Our choice of likelihood function
is predicated on the choice of model $M$.
We use the functions explained in \citet{oayda2024},
where,
for some scalar signal function $f$ at
pixel $i$ described by the unit vector $\mathbf{\hat{p}_i}$,
the likelihood is
\begin{equation}
    \ln \mathcal{L} =
        \sum_{i=1}^{n_{\text{pix}}} \mathcal{N}_i
        \ln \left( \frac{%
                    f(\mathbf{\hat{p}}_i)}{F}
            \right).
        \label{eq:pbp-likelihood}
\end{equation}
Here, $F = \sum_{i=1}^{n_{\text{pix}}} f(\mathbf{\hat{p}}_i)$
is a normalisation term,
summing the function over all unmasked pixels $n_{\text{pix.}}$.
$f$ is the model-dependent term,
whereas the actual value of $\mathcal{N}_i$ and $n_{\text{pix.}}$
depends on the catalogue permutation chosen
(see Section~\ref{subs:permutations}).
The models tested and their associated signal $f$ are as follows:
\begin{itemize}
    \item $M_0$: Monopole, where $f = 1$.
    \item $M_1$: Dipole, where $f = 1 + \mathcal{D} \cos \theta_i$.
    \item $M_2$: Quadrupole, where
        $f = 1 + \mathcal{Q} (\hat{Q}_{jk} \hat{p}_j \hat{p}_k)$
    \item $M_3$: Dipole and quadrupole, where
        $f = 1 + \mathcal{D} \cos \theta_i + 
                    \mathcal{Q} (\hat{Q}_{jk} \hat{p}_j \hat{p}_k)$.
    \item $M_4$: Dipole and octupole, where
            $f = 1 + \mathcal{D} \cos \theta_i
                + \mathscr{O} (\hat{O}_{jkl} \hat{p}_j \hat{p}_k \hat{p}_l)$
\end{itemize}
With respect to our prior likelihoods, we sample from the following
distributions:
\begin{itemize}
    \item $\mathcal{D} \sim \mathcal{U}(0,0.1)$.
    \item $\mathcal{Q} \sim \mathcal{U}(0,0.2)$.
    \item $\mathscr{O} \sim \mathcal{U}(0,0.3)$.
    \item $\mathcal{\phi} \sim \mathcal{U}(0, 2\pi)$.
    \item $\mathcal{\theta} \sim \cos^{-1}(1 - 2u)$ for $u \sim \mathcal{U}(0,0.1)$.
\end{itemize}
Here, $\mathcal{U}(a, b)$ denotes a uniform distribution between $a$ and $b$,
and $\phi$ and $\theta$ represents the azimuthal and polar angles respectively
in radians and equatorial coordinates. Thus, $\phi \in [0, 2\pi)$
and $\theta \in [0, \pi]$.
We chose these prior functions for the dipole direction
to ensure that points are chosen uniformly over the sphere,
taking into account how the area element changes with polar angle.
This enforces all directions to be equally weighted.
Similarly, uniform distributions on the amplitudes are
an expression of the principle of indifference, whereby we favour
no amplitude more than the other.
The increasing domain of $\mathcal{U}$ as we move from dipole
to quadrupole to octupole captures the fact that, based on our
definition of the higher order harmonics, the scale of the
fluctuations become smaller for the same multipole amplitude.

\subsubsection{Model Comparison}
In Bayesian inference,
the model odds ratio for two models $M_i$ and $M_j$ is
\begin{equation}
    O_{ij} = \frac{P(M_i | \mathbf{D})}{P(M_j | \mathbf{D})}
           = \frac{\pi(M_i)}{\pi(M_j)}
             \frac{\mathcal{L}(\mathbf{D} | M_i)}
                {\mathcal{L}(\mathbf{D} | M_j)}
            = \pi_{ij} B_{ij}.
\end{equation}
$\pi_{ij}$ is the prior odds ratio,
representing our a priori beliefs about the relative strength
of each model. Meanwhile, $B_{ij}$ is the Bayes factor -- the ratio
of model marginal likelihoods.
This ratio gives an a posteriori statement about the
relative predictive powers of the two models.
In this study,
we report the natural logarithm of the Bayes factor
defined with respect to the null hypothesis
(the marginal likelihood for model $M_0$).
For example,
we report the strength of model $M_i$ with the metric
\begin{equation}
    \ln B_{i0} = \ln \mathcal{Z}_i - \ln \mathcal{Z}_0.
\end{equation}
We then use Jeffreys's scale, as given in \citet{kass1995},
to convert this numerical scale into a qualitative or intuitive
judgement.

To sample the posterior distribution and compute the marginal likelihood, we use
the \textsc{python} package \textsc{dynesty} \citep{dynesty-v2.1.3}.\footnote{\url{https://pypi.org/project/dynesty/}}
\textsc{dynesty} is an implementation of the Nested Sampling (NS) algorithm, which
samples the posterior in shells of increasing likelihood \citep{skilling2004, skilling2006}.

\section{Results}
\label{sec:results}
In Sections~\ref{sub:dipole_sample}, \ref{sub:quad_sample},
\ref{sub:dip_quad_sample} and \ref{sub:dip_oct_sample}, we describe our inferences
for the pure dipole, pure quadrupole, dipole \& quadrupole and dipole \& octupole samples.
In the following Section at \ref{sub:partial_skies},
we consider the interplay between a survey's fraction of visible sky and source count
and its effect on the inferred dipole parameters.
Lastly, in Section~\ref{sub:effect_priors}, we leave on a cautionary statement
about the effect of one's choice of priors.

\subsection{$S_1$: Dipole Sample}
\label{sub:dipole_sample}
\subsubsection{Effect of mask and source count}
\label{subs:effect_mask_count}
First, for the pure dipole sample $S_1$,
we describe the effect of source count and mask choice on our inferences.
These inferences are based on two pieces of information: 
the posterior distribution for the dipole parameters,
as well as the Bayesian evidence for the dipole model $(M_1)$
compared with that of the monopole null hypothesis $(M_0)$.

\begin{figure}
    \includegraphics[width=0.48\textwidth]{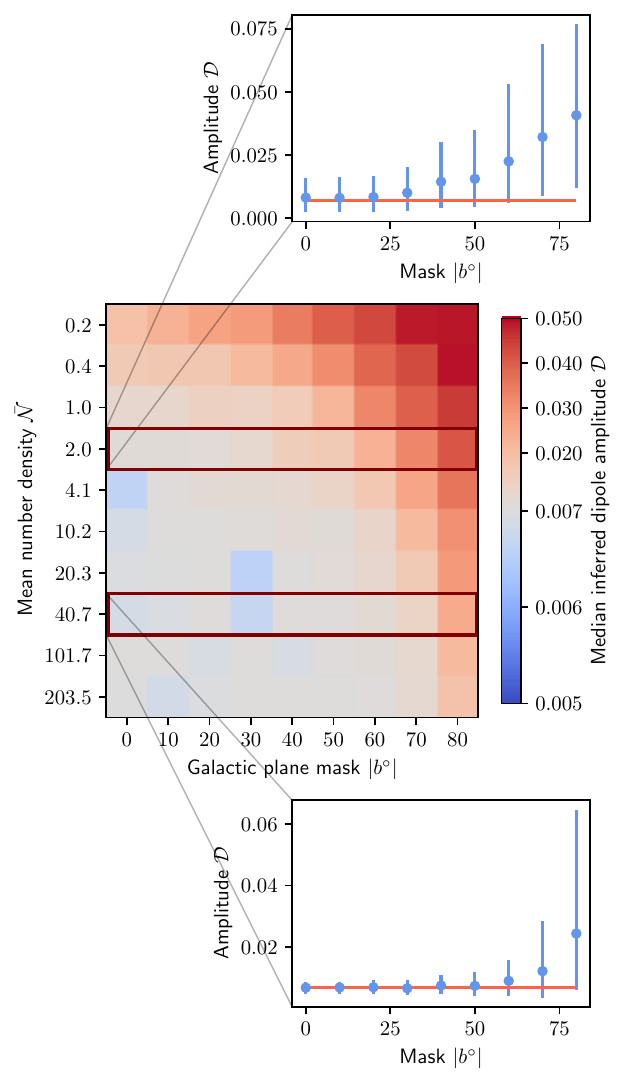}
    \caption{
    Inferred dipole amplitudes in Sample $S_1$ (pure dipole)
    by mask and source density over 50 iterations.
    The `median inferred dipole amplitude' means the median
    of all samples of the dipole amplitude margnal posterior
    at a given permutation (row and column), as described in the main text.
    The true amplitude is $\mathcal{D} = 0.007$, which is grey
    in the colour map used. Red means the inferred amplitude is too high,
    and blue means it is too low.
    The panels above and below the heatmap plot how the inferred
    amplitude changes with choice of mask at $\bar{N} = 2.0$
    and $\bar{N} = 40.7$ respectively.
    \label{fig:dipole_mask_heatmaps_D}
    }
\end{figure}

Our findings relating to the dipole amplitude parameter $\mathcal{D}$ are
represented as a heatmap in Fig.~\ref{fig:dipole_mask_heatmaps_D}.
\begin{figure}
    \includegraphics[width=\linewidth]{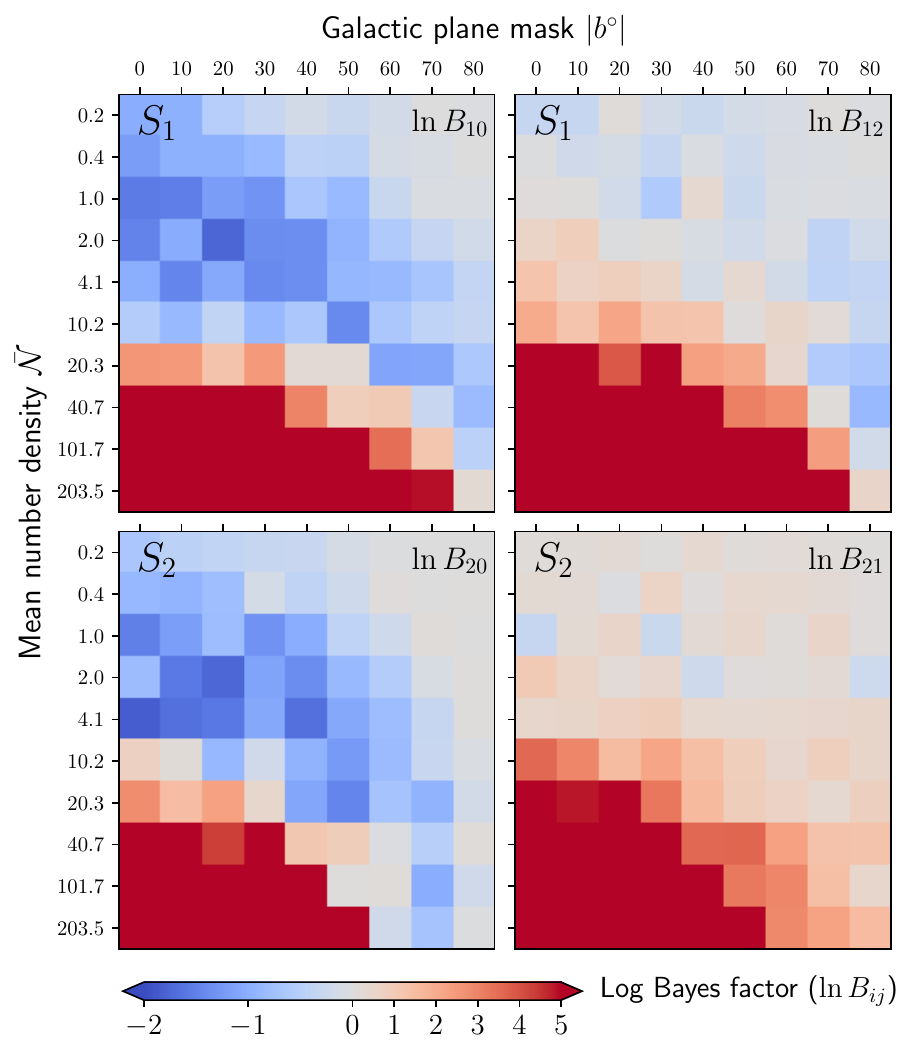}
    \caption{Heatmaps of the log Bayes factor, averaged over many iterations,
    for different models and samples.
    The colour code chosen is intended to reflect Jeffreys's scale.
    Red corresponds to $\ln B_{ij} = 5$, or overwhelming evidence
    for model $M_i$ over a $M_j$.
    Grey corresponds to $\ln B_{ij} = 0$, meaning
    the odds ratio for both models is 1.
    Blue indicates support for $M_j$ over $M_i$.
    \textit{Top left:} dipole versus a monopole ($\ln B_{10}$), dipole sample $S_1$.
    \textit{Top right:} dipole versus a quadrupole ($\ln B_{12}$), dipole sample $S_1$.
    \textit{Bottom left:} quadrupole versus a monopole ($\ln B_{20}$), quadrupole sample $S_2$.
    \textit{Bottom right:} quadrupole versus a dipole ($\ln B_{21}$), quadrupole sample $S_2$.
    \label{fig:dipole_mask_heatmaps_B}
    }
\end{figure}
To produce Fig.~\ref{fig:dipole_mask_heatmaps_D},
once we determined a marginal posterior distribution for $\mathcal{D}$
at a given mask, mean number density and experiment iteration,
we generated 10,000 samples of the dipole amplitude from that distribution.
We repeated this for each experiment iteration (i.e. we generated a total
of 500,000 samples across 50 experiment iterations).
We then computed a $1\sigma$ credible interval (CI)
on the consolidated distribution of all dipole amplitude samples.
The colour bar in Fig.~\ref{fig:dipole_mask_heatmaps_D} represents the median
dipole amplitude; meanwhile, separate panels indicate how the inferred
amplitude changes as a function of Galactic plane mask
at a fixed number density.
The error bars in these panels indicate the limits of the $1\sigma$ CI.
At each iteration, we also computed the log Bayes factor $\ln B_{10}$.
We represent this in the top left pane of
Fig.~\ref{fig:dipole_mask_heatmaps_B} with another heatmap,
where the log Bayes factors for each cell have been averaged over the 50 iterations.

How do we interpret Fig.~\ref{fig:dipole_mask_heatmaps_D} and
Fig.~\ref{fig:dipole_mask_heatmaps_B} (top left) together?
We find that for samples with a sufficient source count,
the median amplitude is consistent with the truth of $\mathcal{D}=0.007$.
This can be seen from the colour scale -- the grey region
indicates that the inferred value is $\approx 0.007$.
However, as the number of sources diminishes -- either due to a low source density
or large mask angle -- the amplitude increases well above the truth,
illustrated by the red region in
the top right of the central pane of Fig.~\ref{fig:dipole_mask_heatmaps_D}.
This is also clear from the two expanded panels above and below the heatmap
at values of large $|b^\circ|$.

\begin{figure}
    \centering
    \includegraphics[width=\linewidth]{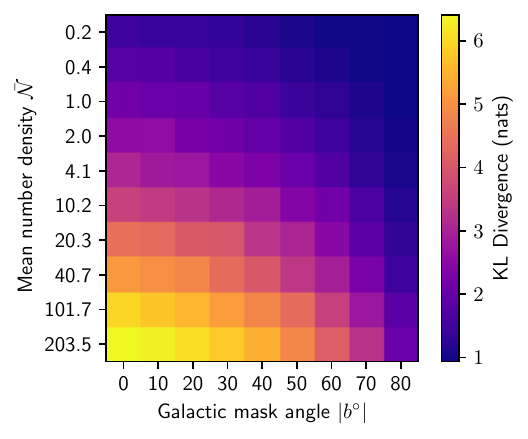}
    \caption{Kulback-Leibler divergence between the posterior for model $M_1$
    and our priors, $D_{\text{KL}} (P \| \pi)$, at different mean number densities and
    Galactic mask angles used in sample $S_1$. The results have been averaged
    over $\approx 20$ iterations. The colour scale indicates
    the divergence in nats, with yellow denoting more information and purple
    less information. They key point is that as the number of sources increases,
    the data offers more information and so the KL divergence between the posterior
    and prior is larger.}
    \label{fig:kl_divs}
\end{figure}

There is a natural way to interpret this trend.
Where the information content of the data is low,
the posterior is expected to more closely align with the prior likelihood function;
our original beliefs have not been substantially tempered by the arrival of new data.
We can quantify this with the Kullback-Leibler divergence
(KL divergence, or relative entropy), where
\begin{equation}
    D_{\text{KL}} (P \| \pi ) =
        \int_{\Omega_{\mathbf{\Theta}}} P(\mathbf{\Theta}) \log
        \frac{P(\mathbf{\Theta})}{\pi(\mathbf{\Theta})} \, d\mathbf{\Theta}
        \label{eq:kl_div}
\end{equation}
for posterior $P(\mathbf{\Theta})$ and prior $\pi(\mathbf{\Theta})$,
as in \eqref{eq:bayes-theorem}, and parameter domain $\Omega_{\mathbf{\Theta}}$.
The KL divergence is the Shannon information averaged over the posterior,
here in nats, and quantifies how much information
is provided by the data $\mathbf{D}$ \citep{handley2019a,handley2019b}.
We therefore computed the KL divergence between our posterior for model $M_1$
and our priors (see Section~\ref{subs:param_optim}), evaluating the integral
numerically after smoothing the distributions from our NS runs with a Gaussian kernel.
These $D_{\text{KL}}$ values, averaged over each iteration, are shown in Fig.~\ref{fig:kl_divs}.
A gradient from purple to yellow can be seen moving from top right to bottom left. This visually confirms that as $N$ increases, the
data becomes more information-rich. For example, at a fixed number density,
masking larger regions decreases the source count, and so $D_{\text{KL}}$ decreases.
The key point is that the factor materially impacting the data's information content
is the number of sources, and not the number density
(which is somewhat arbitrary insofar that one can pick any value for $N_\text{side}$).
There is some subtlety to this statement, however, which we will touch upon later
in Section~\ref{sub:partial_skies}.

With this established, the fact the median dipole amplitude appears to increase
for small $\bar{\mathcal{N}}$ and large mask angle $g_{\text{mask}}^\circ$
is because the data is less informative, and so the original choice of prior
weighs out. Since we used a uniform distribution on $[0, 0.1]$,
it is no surprise that the inferred amplitude tends to $\mathcal{D} =0.05$,
the median of this interval.
This does not mean our inferences are misled with low sample statistics.
One also needs to take into account the Bayes factors,
again as shown in the top left of Fig.~\ref{fig:dipole_mask_heatmaps_B}.
Where the inferred amplitude tends to be discrepant (too large),
the Bayes factors either give slightly stronger evidence for a monopole
(blue regions) or prefer no model over the other (grey regions).
This tells us that we cannot take
the inferred amplitude at face value, since the dipole model
does not offer great explanatory power -- or even gives marginally worse
predictions of the data than a monopole while balancing model complexity.

\begin{figure}
    \centering
    \includegraphics[width=\linewidth]{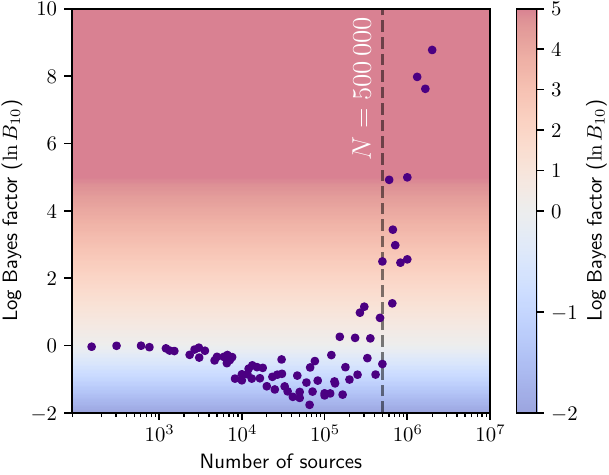}
    \caption{$\ln B_{10}$ (dipole vs. monopole) as a function of the number of sources $N$
            in the synthetic sample. The colour scale also indicates the log Bayes factor,
            matching that used in Fig.~\ref{fig:dipole_mask_heatmaps_B}. The dashed line
            corresponds to $N = 500\,000$.}
    \label{fig:nsource-b10_S1}
\end{figure}

For what source counts can the dipole be accurately inferred and recovered?
To see this, in Fig.~\ref{fig:nsource-b10_S1},
we plot how our log Bayes factors change
as a function of source count at each catalogue permutation.
By $N \approx 10^{6}$ sources,
$\ln B_{10} \gtrapprox 5$, suggesting overwhelming support for a dipole
over a monopole \citep{kass1995}.
Also, there is generally positive support for a dipole
beginning at $N \approx 5 \times 10^{5}$ (dashed line).
One curious feature is that for $N \leq 10^3$,
the log Bayes factor is about zero, while around $N=10^4$ to $N=10^5$,
it dips to its lowest extent near $-2$ before increasing again for $N>10^5$.
We believe this arises from there being insufficient data (small $N$)
to make any statement about the models,
whereas for the mid-range values the Occam factor dominates.
The Occam factor, implicit in the marginal likelihood, penalises model complexity.
This means that the dipole with three model parameters
suffers a greater penalty than the simpler (though inaccurate)
zero-parameter monopole model.
Even so, the effect of the penalty is not particularly severe; there is only very slim
to equal preference for a monopole compared to a dipole.
But once this valley of indifference is crossed,
one climbs the slope of knowledge as the number of sources
increases past $10^5$. 

\begin{figure*}
    \centering
    \includegraphics[width=0.9\linewidth]{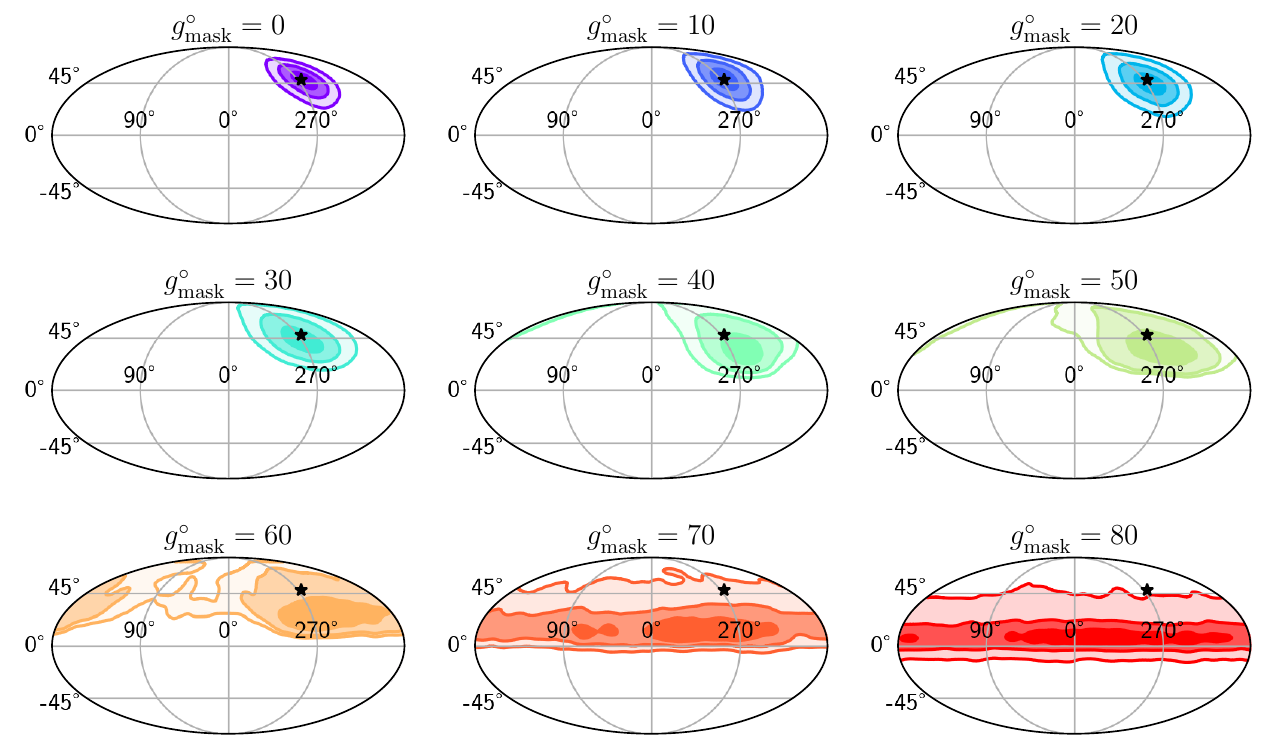}
    \caption{Evolution of the distribution of dipole directions with Galactic plane
    mask $g_{\text{mask}}^\circ$ and $\bar{N} = 40.7$ (Mollweide projection),
    consolidated over 50 iterations for each mask. The black star indicates
    the direction of the CMB dipole. The contours enclose $0.5\sigma$ levels
    of posterior density.}
    \label{fig:bayessky_proj_mask}
\end{figure*}

\begin{figure*}
    \centering
    \includegraphics[width=0.8\linewidth]{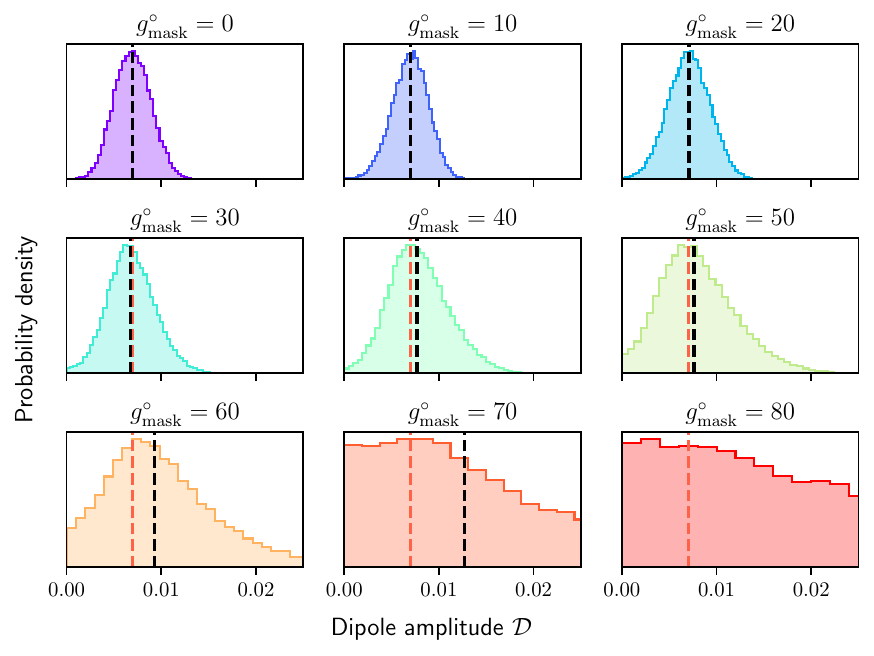}
    \caption{As for Fig.~\ref{fig:bayessky_proj_mask} but for the distribution
        of dipole amplitudes. The colours used here and in Fig.~\ref{fig:bayessky_proj_mask} match for the same Galactic plane mask.
        The black dashed lines indicates the median of the distribution,
        whereas the red dashed line is the sample truth $\mathcal{D} = 0.007$.}
    \label{fig:bayes_amp_mask}
\end{figure*}

To show the full posterior instead of only summary statistics for $\mathcal{D}$,
we isolate the set of simulations performed with samples
of mean density $\bar{N} = 40.7$ and plot how the inferred dipole
parameters change with choice of Galactic plane mask $g_{\text{mask}}^\circ$.
These distributions have been consolidated over the 50 runs, as described earlier,
and are shown in Fig.~\ref{fig:bayessky_proj_mask}
and Fig.~\ref{fig:bayes_amp_mask}.
The former figure projects the joint distribution
for the dipole direction $(l,b)$ onto the celestial sphere,
and the latter shows the PDF explicitly for $\mathcal{D}$.
In Fig.~\ref{fig:bayessky_proj_mask},
one can see that for lower masks ($0^\circ$ to $30^\circ$),
the direction is broadly consistent with the truth.
As the masking angle increases,
the distribution shifts to lower Galactic latitudes and begins to widen
around the Galactic plane,
although the uncertainties increase dramatically.
Meanwhile, in Fig.~\ref{fig:bayes_amp_mask},
the dispersion of the dipole amplitude increases
such that near-zero amplitudes become increasingly more probable.
Again, this has to be taken together with the Bayes factors,
which tell us that for $g_{\text{mask}}^\circ \geq 50$
with $\bar{\mathcal{N}} = 40.7$,
the dipole and monopole models are more or less on equal footing.
In other words, 
where the dipole direction and amplitude estimates tend to be less accurate,
one would conclude that there is insufficient data or information to support
either model.
This is as opposed to giving an overly-confident estimate of the dipole amplitude
and/or direction.

\subsubsection{Fitting a dipole vs. quadrupole}
So far we have only been concerned with fitting the dipole ($M_1$).
What happens if we try to fit a quadrupole ($M_2$) on our pure dipole sample $S_1$?
In a separate set of simulations,
we verified that a quadrupole -- as defined in Section~\ref{sub:stats-regime} --
does not have anomalously more support compared to a dipole.
This is confirmed in the top right pane of Fig.~\ref{fig:dipole_mask_heatmaps_B}
where we use the metric $\ln B_{12}$.
Apart from very minor edge cases at $|b^\circ| = 80$ and low source densities,
the dipole dominates -- unless the data has insufficient information content,
as was touched on above.

\subsection{$S_2$: Quadrupole Sample}
\label{sub:quad_sample}
Turning to our simulated quadrupole sample ($S_2$),
we present the heatmap of Bayes factors
in the bottom row of Fig.~\ref{fig:dipole_mask_heatmaps_B}.
For this sample,
the quadrupole becomes the overwhelmingly dominant model with source counts
in excess of $\approx 1$ million.
For lower counts,
either the monopole is marginally favoured,
or all models are approximately on equal footing.
These results suggest that one needs more sources to reach high levels
of support for an intrinsic quadrupole compared to a dipole. 
Also, even with heavily masked skies and/or low
source counts, an intrinsic quadrupole is never mistaken for a dipole.
If this were the case, the dipole would have a higher marginal likelihood,
which would be indicated by blue regions in the bottom right pane of Fig.~\ref{fig:dipole_mask_heatmaps_B}.
We discuss the nature of the quadrupole posterior distribution
in the following section.

\subsection{$S_3$: Dipole and Quadrupole Sample}
\label{sub:dip_quad_sample}
\begin{figure*}
    \centering
    \includegraphics[width=0.85\linewidth]{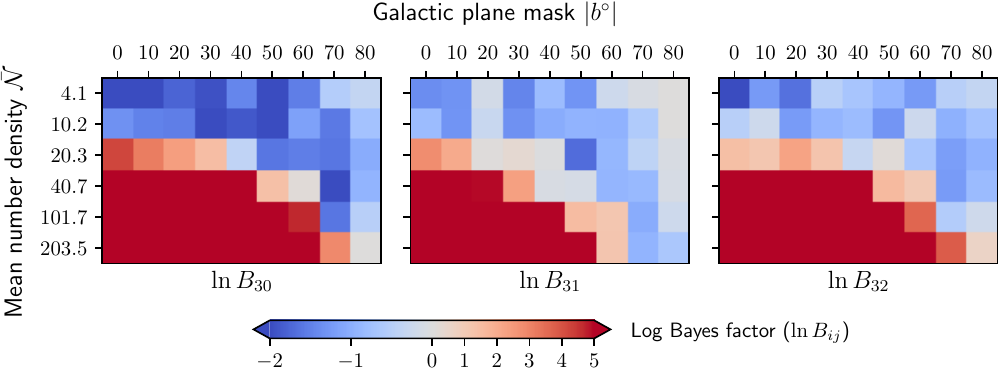}
    \caption{As for Figures~\ref{fig:dipole_mask_heatmaps_B}
    except with sample $S_3$, and, from left to right,
    the dipole \& quadrupole model ($M_3$) is compared to the monopole,
    dipole and quadrupole models ($M_0$, $M_1$, $M_3$ respectively).
    \textit{Left:} $M_3$ vs $M_0$.
    \textit{Middle:} $M_3$ vs $M_1$.
    \textit{Right:} $M_3$ vs $M_2$.}
    \label{fig:bij_s4}
\end{figure*}

We first give our Bayes factors for the dipole \& quadrupole sample ($S_3$)
in Fig.~\ref{fig:bij_s4}.
These have been separated from Fig.~\ref{fig:dipole_mask_heatmaps_B} purely
for aesthetic reasons.
Also, the rows with number densities less than 4.1 have been dropped since,
as based on the foregoing results, the data will not be sufficiently
information rich to provide strong support for one model over the other.
One can see from a cursory inspection of Fig.~\ref{fig:bij_s4}
that the dipole and quadrupole model ($M_3$) consistently dominates ($\ln B_{ij} \geq 5$)
over the other models near the bottom left region of the heatmaps.
Using the middle panel as the limiting factor ($\ln B_{31}$) and again referring
to the source count map in Fig.~\ref{fig:expected_count_map},
this corresponds to a number of sources $N \approx 1.5$ million.

\begin{figure*}
    \centering
    \includegraphics[width=\linewidth]{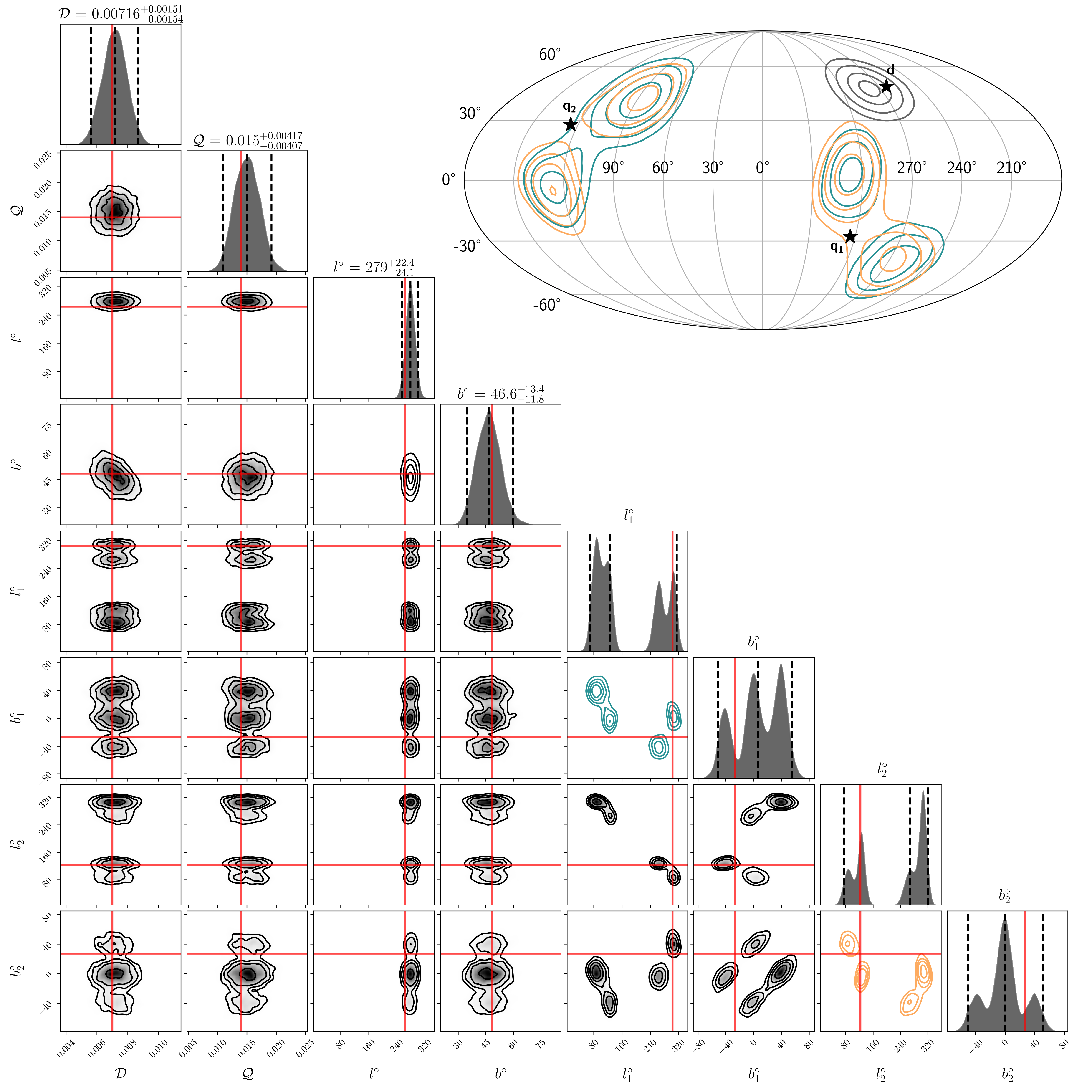}
    \caption{Posterior for model $M_3$ after one run at $\bar{\mathcal{N}}=203.5$
    and $g_{\text{mask}}^\circ = 30$ ($N\approx 5$ million)
    for sample $S_3$. $\mathcal{D}$ and $\mathcal{Q}$
    are the dipole and quadrupole amplitudes respectively, whereas $l$, $b$, $l_1$, $b_1$,
    $l_2$, $b_2$ specify the directions of the dipole vector $\mathbf{d}$
    and the two quadrupole unit vectors $\hat{\mathbf{q}}_1$ and $\hat{\mathbf{q}}_2$ respectively.
    \textit{Corner plot; bottom left:} The quoted confidence intervals (where applicable) give a
        $2\sigma$ level of statistical significance. The quadrupole directions are left without
        intervals due to the more complex (multi-modal) structure of their marginal posteriors.
    \textit{Projection; top right:} The distributions for the dipole and quadrupole directions
        are projected onto the sky. The contours give $0.5\sigma$ levels of posterior density.
        The yellow contours denote the joint distribution for $(l_2, b_2)$, whereas the teal
        contours give the distribution for $(l_1, b_1)$.}
    \label{fig:dq_corner}
\end{figure*}

How does the posterior for $M_3$ look when it is the preeminent model?
To see this, we reproduce the results from one iteration at
$\bar{\mathcal{N}}=203.5$ and $g_\text{mask}^\circ = 30$ (Fig.~\ref{fig:dq_corner}).
There, ($l, b$) denotes the dipole direction in Galactic coordinates,
whereas $(l_1, b_1)$ and $(l_2, b_2)$ denote the directions of the
two quadrupole unit vectors.
In the top right of the same figure, the posterior probability distribution for the
dipole and quadrupole directions are projected onto the sky in Galactic coordinates.
The teal and yellow colours guide the eye in matching the joint direction distributions
in the corner plots to the sky projection. 

Now, while the amplitudes of the dipole $\mathcal{D}$ and quadrupole $\mathcal{Q}$
are recovered well,
there is evidently multi-modal structure in the direction posterior.
This warrants an explanation.
Recall that the true directions for the quadrupole unit vectors in Sample $S_3$
point towards $(l_1,b_1) = (302\dotdeg9, -27\dotdeg1)$
and $(l_2,b_2=122\dotdeg9, 27\dotdeg1)$.
If we denote these vectors as $\hat{\mathbf{q}}_1$
and $\hat{\mathbf{q}}_2$, then since the priors on the two unit vector
positions sample the entire celestial sphere,
it is unsurprising that there should be at least
two peaks in the marginal distributions: one for ($l_1$, $b_1$), and one for ($l_2$, $b_2$).
That is, in sampling for $\hat{\mathbf{q}}_1$, a peak corresponding to
$\hat{\mathbf{q}}_2$ also appears.
However, there are additional peaks apparent in the sky projection of Fig.~\ref{fig:dq_corner}.
Namely, there are four maxima per quadrupole direction.
This arises because of ambiguity about the sign of the vector.
If we take $\hat{\mathbf{q}}_1 \otimes \hat{\mathbf{q}}_2$ to construct
the quadrupole tensor,
then $(-\hat{\mathbf{q}}_1) \otimes (-\hat{\mathbf{q}}_2)$
produces the same tensor.
Thus, when we sample the underlying probability distribution,
each step in the chain corresponds to either the pair
$\hat{\mathbf{q}}_1$ and $\hat{\mathbf{q}}_2$
or $-\hat{\mathbf{q}}_1$ and $-\hat{\mathbf{q}}_2$.
Since both pairs of unit vectors produce the same quadrupole signal,
we should see two sets of two peaks in posterior space: each peak being identical.
This explains the four peaks in the top right of Fig.~\ref{fig:dq_corner}.
Any difference in their shape arises from the coarseness of our numerical estimation, which we consider further in Section~\ref{sub:dip_oct_sample}.

This being said, the takeaway message is that --
even where the sky is substantially masked --
we can disentangle the individual contributions from the dipole and quadrupole
in sample $S_3$.
Note in Fig.~\ref{fig:dq_corner} that the marginal posterior distributions
for the dipole amplitudes are well-constrained, and the truths lie
close to the medians of each distribution.
This is important, since it suggests we are not plagued by any mode mixing
effect, at least between $\ell=1$ and $\ell=2$.

\subsection{$S_4$: Dipole and Octupole Sample}
\label{sub:dip_oct_sample}
Since computational complexity is more of an issue when fitting an octupole,
we restrict our analysis to one set of sample parameters:
$\bar{\mathcal{N}}=4000$ and $g_\text{mask}^\circ = 30$, yielding $N\approx98$ million.
Of course, this is a large number of sources in comparison to what is feasible
in current surveys (but not in the future, see Section~\ref{sub:future}).
Nonetheless, half the sky has been thrown out,
and we are mainly interested in whether or not the dipole and octupole
parameters can be accurately recovered.
In other words, is there crosstalk or leakage between the two multipoles,
as is claimed by \citet{abghari2024} to be an issue for CatWISE2020?

\begin{figure*}
    \centering
    \includegraphics[width=\linewidth]{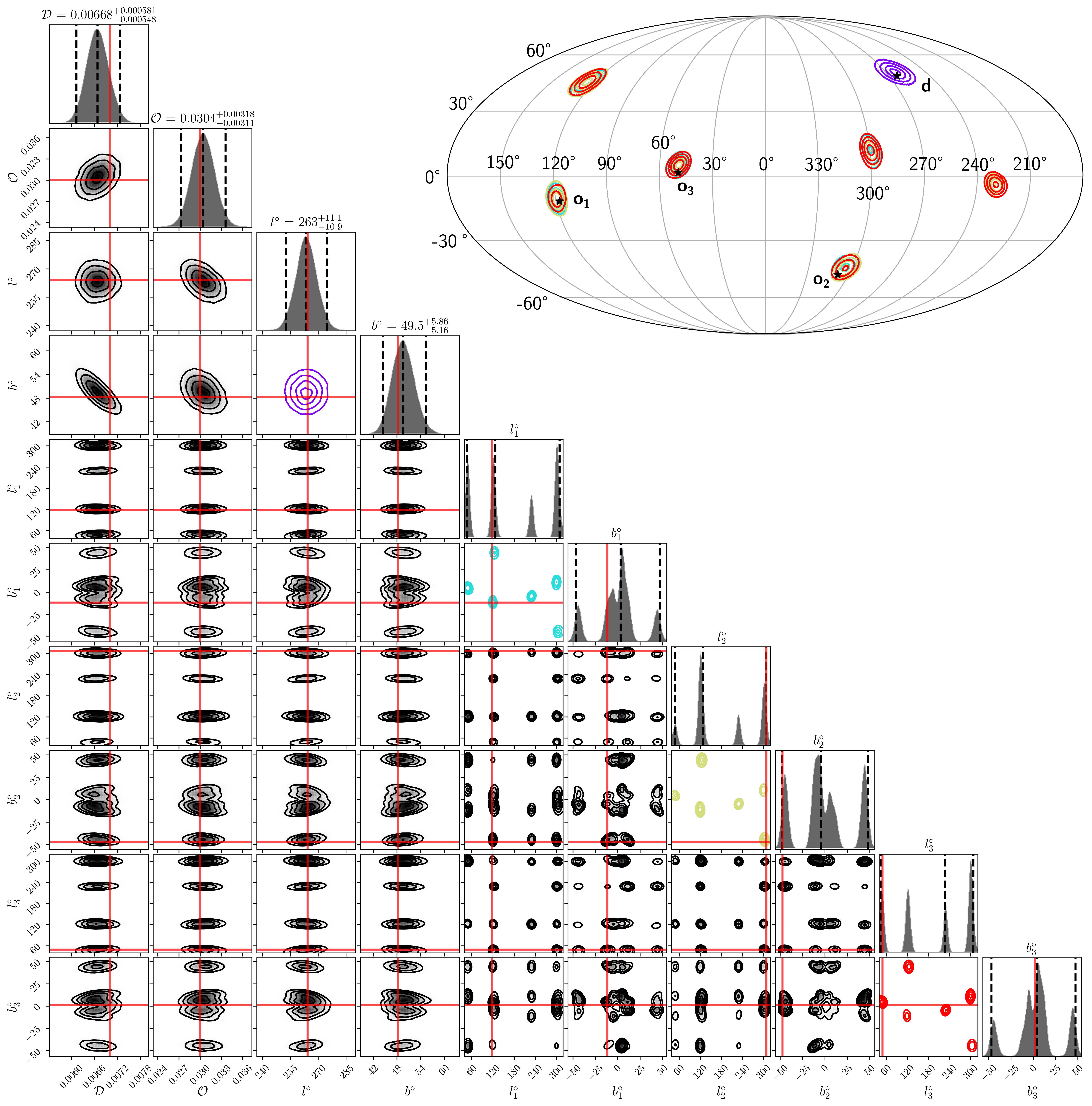}
    \caption{Posterior for model $M_4$ with $N\approx 98$ million
    and $g^\circ_{\text{mask}} = 30$ for Sample $S_4$.
    The details of the plot are the same
    as Fig.~\ref{fig:dq_corner}, except now the vectors $\mathbf{o_1}$, $\mathbf{o_2}$
    and $\mathbf{o_3}$ represent the three octupole unit vectors.}
    \label{fig:do-corner}
\end{figure*}

One issue that emerges for higher order multipoles is the
posterior becoming increasingly multi-modal.
For the quadrupole, each joint $(l,b)$ distribution has four peaks; for the octupole,
this increases to six.
In the NS algorithm, there is the risk of mode `die-off': an irreversible process in which
all of the live points move off of one mode to another, preventing that original mode
from being properly sampled.
In attempt to safeguard against this,
we used a high number of live points ($n_{\text{live}} = 40\,000$).
However, this increases the time taken for each NS run,
with the time complexity scaling linearly with $n_{\text{live}}$ \citep{buchner2023}.
More pertinently, we saw scarcely any improvement in the shape of the modes
with a larger number of live points.
We therefore switched from \textsc{dynesty} to the package 
\textsc{UltraNest}\footnote{\url{https://johannesbuchner.github.io/UltraNest/}}
\citep{ultranest}, which implements the nested sampling
Monte Carlo algorithm MLFriends \citep{buchner2016, buchner2019}.
For our purposes, \textsc{UltraNest} appears to be better-equipped at sampling
the multi-modal 10-dimensional posterior of the dipole \& octupole model ($M_4$).
We reproduce the results from one such run in Fig.~\ref{fig:do-corner}.
As can be seen there, each of the true octupole directions (of which there are six,
counting the degeneracies) has roughly the same number of contours in each 2D $(l,b)$ space.
Further, the modes of each angle parameter marginal distribution
are roughly the same height.

After all this, the critical point is that the
dipole and octupole amplitudes are recovered accurately,
even though half the sky has been masked.
Even on incomplete skies, power leakage from the octupole to the dipole
is not occurring.
Thus, as we explained earlier,
our only bottleneck with multipole inference is the information content of the data --
as well as the computational complexity of our procedure.

\subsection{Inference with Partial Skies: Designing a Small-footprint Survey}
\label{sub:partial_skies}
\subsubsection{Continuous surveys}
We mentioned earlier that the number of sources $N$ materially impacts the
information content of the data.
This statement, however, needs to be qualified by the fraction of sky which is
masked in the sample.
Before, we looked at how our inferences change with different Galactic plane cuts.
Now, we look at the case of a single continuous region of visible sky, with the remainder
being masked.

\begin{figure}
    \centering
    \includegraphics[width=\linewidth]{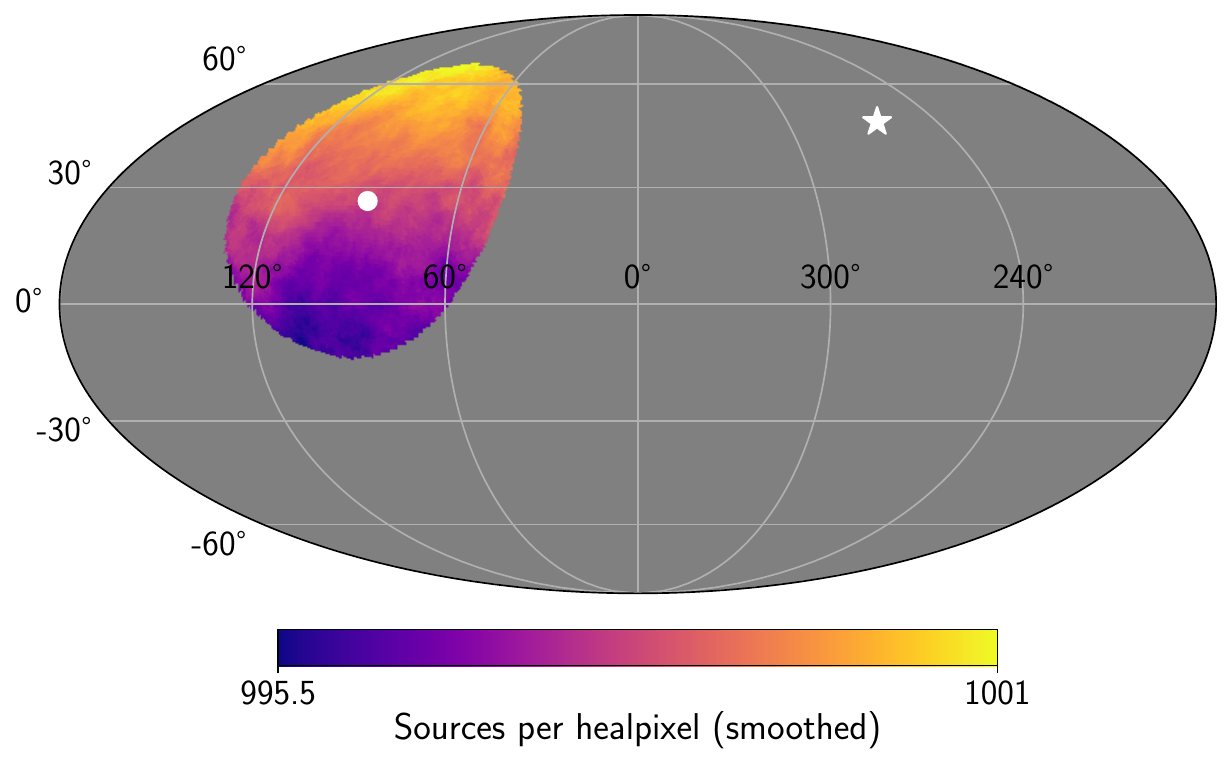}
    \includegraphics[width=\linewidth]{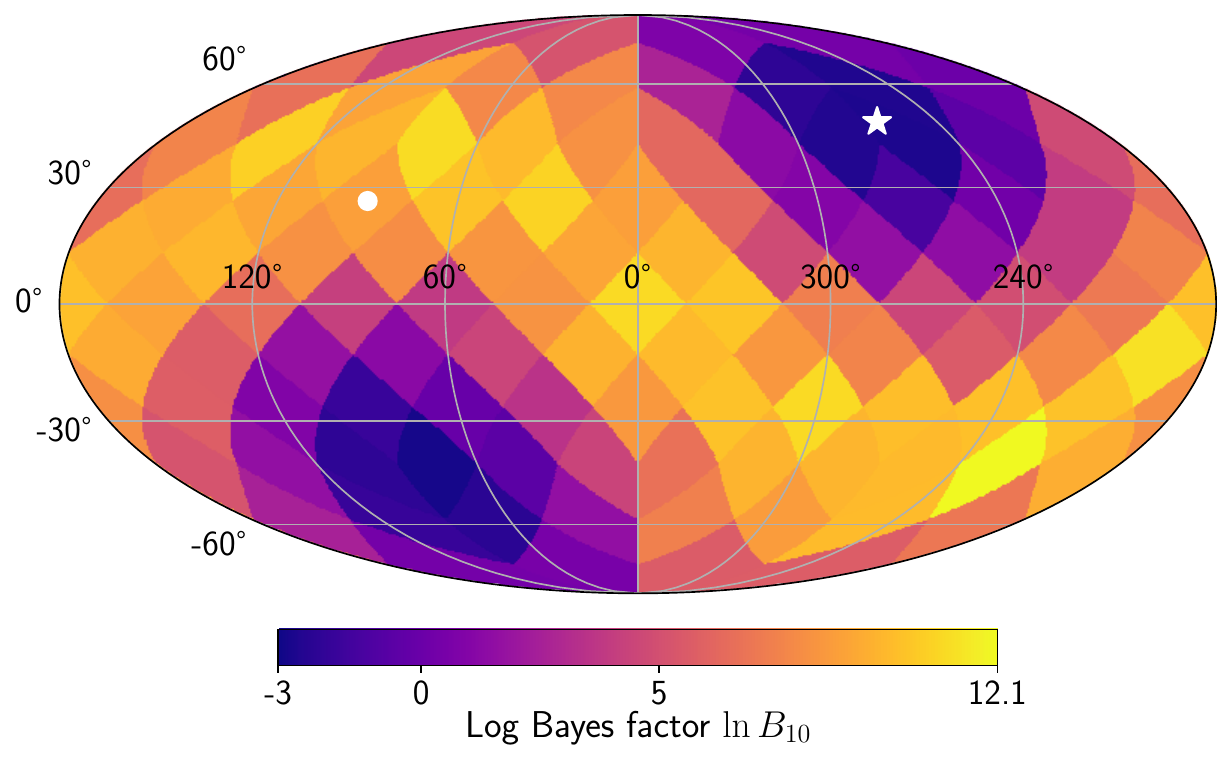}
    \caption{Inferences made by location (Galactic coordinates) of visible
            patch of sky with respect to the direction of the dipole vector
            (white star).
            \textit{Top:} Location of one generated patch of visible sky,
            with masked areas in grey.
            The patch is $40^\circ$ in radius, corresponding
            to a sky fraction of $f_{\text{sky}} \approx 11.7\%$.
            The source count per healpixel ($\bar{\mathcal{N}} = 1000$)
            is given after applying a moving average
            with angular size of 1 steradian.
            \textit{Bottom:} Log Bayes factor (dipole vs. monopole)
            at a given healpixel ($N_\text{side} = 3$),
            defined such that the centre of the unmasked patch
            of sky lies at the centre of the healpixel.
            The white dot guides the eye to the centre of the selected
            sky patch (top row) and the corresponding pixel centre (bottom row).}
    \label{fig:bmap_slices}
\end{figure}

An important consideration is the location of this visible patch of sky with respect
to the underlying dipole signal.
We generally find that if the patch of sky aligns with the dipole maxima or minima,
the Bayes factor $\ln B_{10}$ suggests support for a monopole,
at least for the case of $\bar{\mathcal{N}} = 1000$ and
$\mathcal{D} = 0.007$ in our pure dipole sample.
This effect is exhibited in Fig.~\ref{fig:bmap_slices}.
There, one such patch of sky is indicated in the top pane.
It has a $40^\circ$ radius, so the fraction of visible sky is $f_{\text{sky}} = 11.7\%$
and the slice contains about 5.75 million sources.
We take this slice and slide it across the sky, such that the centre of the slice
(the white dot in the top and bottom panes) lies on the centres of pixels
created from an $N_{\text{side}} = 3$ (108 healpixels) healpy map.
As the centre of the sky patch moves towards the dipole equator
(90$^\circ$ away from the dipole maxima/minima),
the average log Bayes factor is maximised with beyond overwhelming support for a dipole.
This can be seen by the yellow/orange band running through the sky map in
the bottom pane of Fig.~\ref{fig:bmap_slices}.

\begin{figure}
    \centering
    \includegraphics[width=\linewidth]{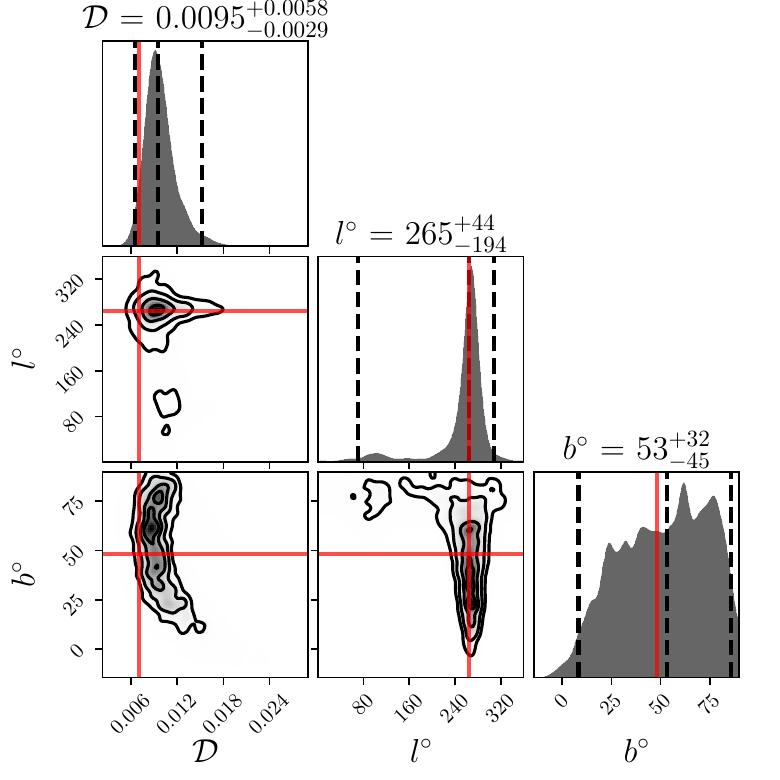}
    
    \vspace{2mm}\hfill\includegraphics[width=0.9\linewidth]{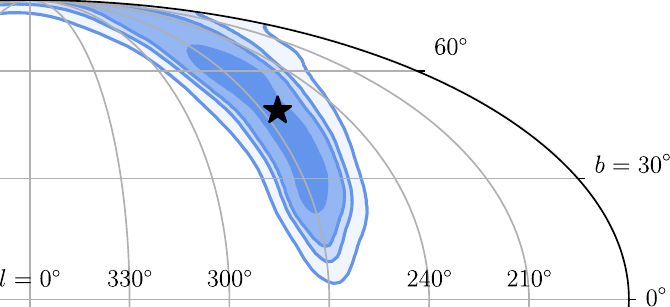}
    \caption{\textit{Top:} Corner plot showing the posterior distribution
    from one run with a slice of $40^\circ$ sky (see Fig.~\ref{fig:bmap_slices})
    near the dipole equator with 5.75 million sources.
    The dashed lines enclose a $2\sigma$ CI,
    while the contours indicate intervals of $0.5\sigma$.
    The red lines indicate the true values in the sample.
    \textit{Bottom:} Projection of joint distribution for $(l^\circ, b^\circ$)
    onto the sky in Galactic coordinates.
    The black star indicates the true dipole direction.
    The contours again indicate intervals of $0.5\sigma$.}
    \label{fig:degen_slice}
\end{figure}

So, if the patch is in an optimal location,
the Bayes factors on net suggest very strong support for a dipole.
But how do the posterior distributions look?
In general, we find that even if one situates the unmasked patch of sky
at the dipole equator, there is a degeneracy in the joint posterior
for the dipole direction. Specifically, while the dipole amplitude
and direction in Galactic longitude are recovered well, the polar angle
is highly degenerate.
An example is given from one run in Fig.~\ref{fig:degen_slice}.
The degeneracy in $b^\circ$ can clearly be discerned; nonetheless,
the result is consistent with the true values given the
uncertainties.

\begin{figure}
    \centering
    \includegraphics[width=\linewidth]{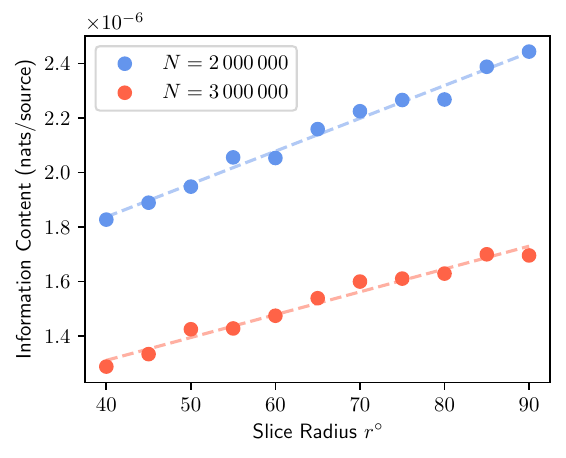}
    \caption{Information content (nats/source; KL divergence divided by
    total number of sources) as a function of slice radius $r^\circ$,
    defined as the angular separation from the centre of the slice
    to its border with the masked region.
    A simple linear fit has been plotted with a dashed line to guide the eye.
    As $r^\circ$ increases, each source carries more information.
    As $N$ increases, each source carries less information,
    though the total information is higher.}
    \label{fig:info-content-r}
\end{figure}

Can the degeneracy be broken while maintaining sparse sky coverage?
In one sense, yes; we could increase $N$ arbitrarily,
adding more information until the distributions localise around the truths.
We confirm this is indeed the case in the Appendix at Fig.~\ref{fig:highN-example},
in which we use a $r = 40^\circ$ slice centred at the southern equatorial pole
containing $\approx 150$ million sources.
However, clearly $N$ is not the only contributing factor to the
information content, with the fraction of visible sky $f_{\text{sky}}$
playing some part.
To see this,
we fix $N = 2 \times 10^6$ and $N = 3 \times 10^6$,
then profile how the KL divergence changes
with the radius of the slice of visible sky.
The slice is again centred at the southern equatorial pole.
As in Fig.~\ref{fig:info-content-r},
we observe a linear dependence of the KL divergence per source (nats/source)
on the slice radius $r^\circ$, at least between 
radii (sky fractions) of 40$^\circ$ (11.7\%) to 90$^\circ$ (50\%).
In other words, at a fixed $N$,
the information per source is higher for
samples covering more of the celestial sphere than those localised
to small regions.

\begin{figure}
    \centering
    \includegraphics[width=\linewidth]{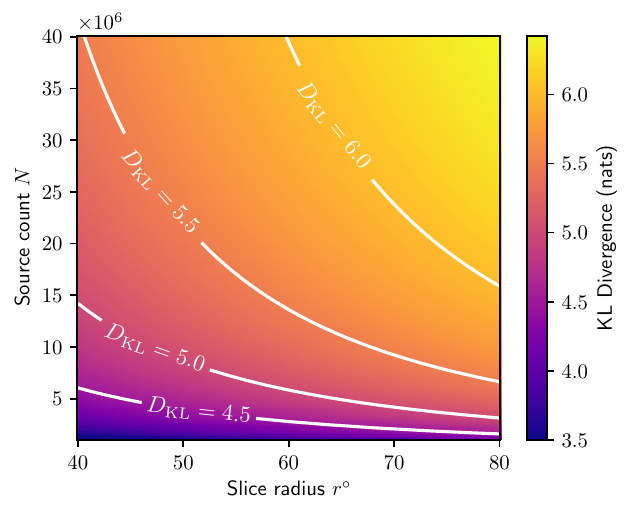}
    \caption{Kl divergence (nats) for different sample source counts
    and slice radii, as determined from a fit to the function
    $D_{\text{KL}} = A \log_{10} N + Br^\circ + C$ with the computed
    values of $D_{\text{KL}}$. Contours denoting lines of equal
    $D_{\text{KL}}$ are labelled in white.}
    \label{fig:inf_matrix}
\end{figure}

In this case, if in some empirical survey we are limited to a fixed number of sources,
maximising $r^\circ$ or $f_{\text{sky}}$ will yield the most information.
But a large survey area is in many cases unachievable, for example owing to
declination limits for ground-based radio surveys
or systematic effects from the Galactic plane \citep[see e.g.][]{secrest2021,mittal2024}.
This begs the question: if we are limited to a small patch of sky in regions
where the survey is maximally-sensitive and/or minimally contaminated by systematic effects,
how many sources are needed to recover dipole parameters
to a requisite degree of confidence?
Continuing with the information-theoretic approach,
we can choose some arbitrary information threshold as a proxy for the size
of the uncertainties (more accurately the change between prior and posterior
distributions), and permit values of $N$ and $r^\circ$ which meet this threshold.
This is something of a heuristic technique,
and importantly the exact value of the threshold will be very sensitive
to the choice of prior distributions and the model parameters (see \eqref{eq:kl_div}).
We can then use the computed values of the KL divergence to create a scalar
function for $D_{\text{KL}} = D_{\text{KL}}(N, r^\circ)$ (Fig.~\ref{fig:inf_matrix}).
By inspection,
we fit $D_{\text{KL}}(N, r^\circ) = A \log_{10} N + (r^\circ)^B - N^C + D$
for some constants $A$, $B$, $C$ and $D$.
This function only serves the purpose for generating
approximate predictions for $D_\text{KL}$ on a continuous range
of $N$ and $r^\circ$ in the domains we tested,
so there is not much significance in its exact functional form.
For example, with a threshold of $D_{\text{KL}} = 5.0$, we would need
$\approx 14$ million sources in a $40^\circ$ slice or $\approx 3$ million
sources in an $80^\circ$ slice.

Our principal concern is breaking the degeneracy in one of the
dipole angle parameters.
Based on an inspection of the typical posteriors
as a function of $D_{\text{KL}}$, we find that moving from a divergence
of 5.0 nats to 5.5 nats is matched with a sizeable shrinkage in the
uncertainties for $b^\circ$.
To reflect this, we compute a $2\sigma$ credible interval for the
marginal distribution of $b^\circ$ (i.e. $[b_{\text{low}}, b_{\text{high}}]$)
where $D_{\text{KL}}=5.0$ and where $D_{\text{KL}}=5.5$,
and then determine $\Delta b = b_{\text{high}} - b_{\text{low}}$.
We find that typical values of $\Delta b$ fall around $42^\circ$ for
the higher KL divergence threshold,
whereas $\Delta b \approx 55^\circ$ for the lower threshold.

Similarly for the dipole amplitude,
to provide a more concrete measure of statistical significance,
we take the marginal posterior for $\mathcal{D}$ where
the chosen values of $N$ and $r^\circ$ yield $D_{\text{KL}}=5.5$,
then compute the probability
$P(\mathcal{D} < 2\times D_{\text{CMB}} = 0.014)$.
Recall that we imprinted onto the sample an intrinsic dipole
with magnitude $\mathcal{D}=0.007$,
which we set to be the `true CMB amplitude'.
Then, in effect, we are computing the probability that the inferred
dipole is less than two times the expectation, a typical value
that has been reported across the literature.
We turn this probability into a level of significance using
a one-sided normal distribution.
At $D_{\text{KL}} = 5.5$, we find a typical significance of
$(4.6 \pm 0.6)\sigma$ for the $40^\circ$ slice, and
$(6.4 \pm 0.8) \sigma$ for the $80^\circ$ slice,
with intermediate values for slices in between these two extremes.
Although we are moving along an iso-information contour,
the significance drops for smaller slice angles largely because the
marginal distribution for $\mathcal{D}$ becomes positively skewed.
Nonetheless,
in either case the significance is $\geq 4 \sigma$.
Meanwhile, with $D_{\text{KL}} = 5.0$, we find
the significance is $(2.5 \pm 0.5)\sigma$ at $40^\circ$
and $(4.3 \pm 0.7)\sigma$ at $80^\circ$.

Again, a spectrum of values for $N$ and $r^\circ$ can be chosen;
the numbers we have quoted here only give some guidance
on the consequences of those choices.
Nonetheless, choosing sample parameters $N$ and $r^\circ$
such that $D_{\text{KL}} = 5.5$ offers a good reduction in the uncertainties
on the dipole direction, as well as a $\geq 4\sigma$ statistical significance
for the dipole being less than twice the expectation.
This threshold corresponds to $N\approx 42$ million with $r^\circ=40$
and $N\approx6.6$ million with $r^\circ=80$.

\begin{figure}
    \centering
    \includegraphics[width=\linewidth]{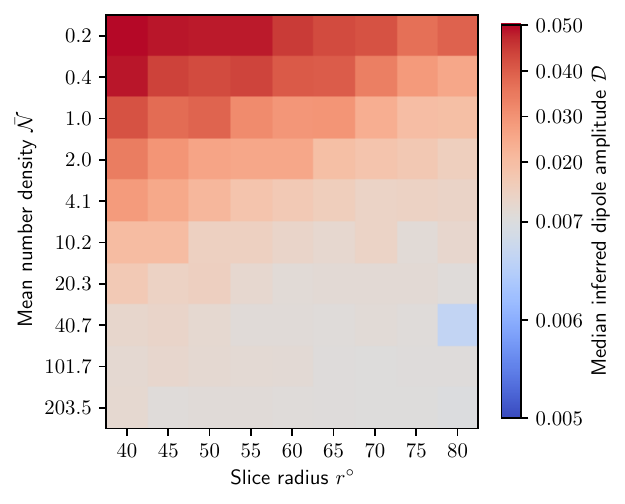}
    \caption{%
        As for Fig.~\ref{fig:dipole_mask_heatmaps_D}
        (using the same pure dipole template $S_1$),
        except the sample consists of
        small slices centred at the south equatorial pole of varying radii.
        An example slice is shown in the top of Fig.~\ref{fig:bmap_slices},
        though note that that one is centred near the north equatorial pole.
    }
    \label{fig:inferred_D_hemi}
\end{figure}

Lastly, in Fig.~\ref{fig:inferred_D_hemi}
we present inferences on the dipole amplitude as was done in Fig.~\ref{fig:dipole_mask_heatmaps_D}
but with the slices we defined above (centred at the southern equatorial pole)
instead of for different galactic masks.
A similar conclusion is reached: as long as
the source counts are sufficient (given some angular breadth),
the median amplitude tends to the truth of $\mathcal{D} = 0.007$.
As a rough estimate,
this is where the number of sources $N \gtrapprox 500\,000$.

\subsubsection{Discontinuous surveys}
\begin{figure}
    \centering
    \includegraphics[width=\linewidth]{
        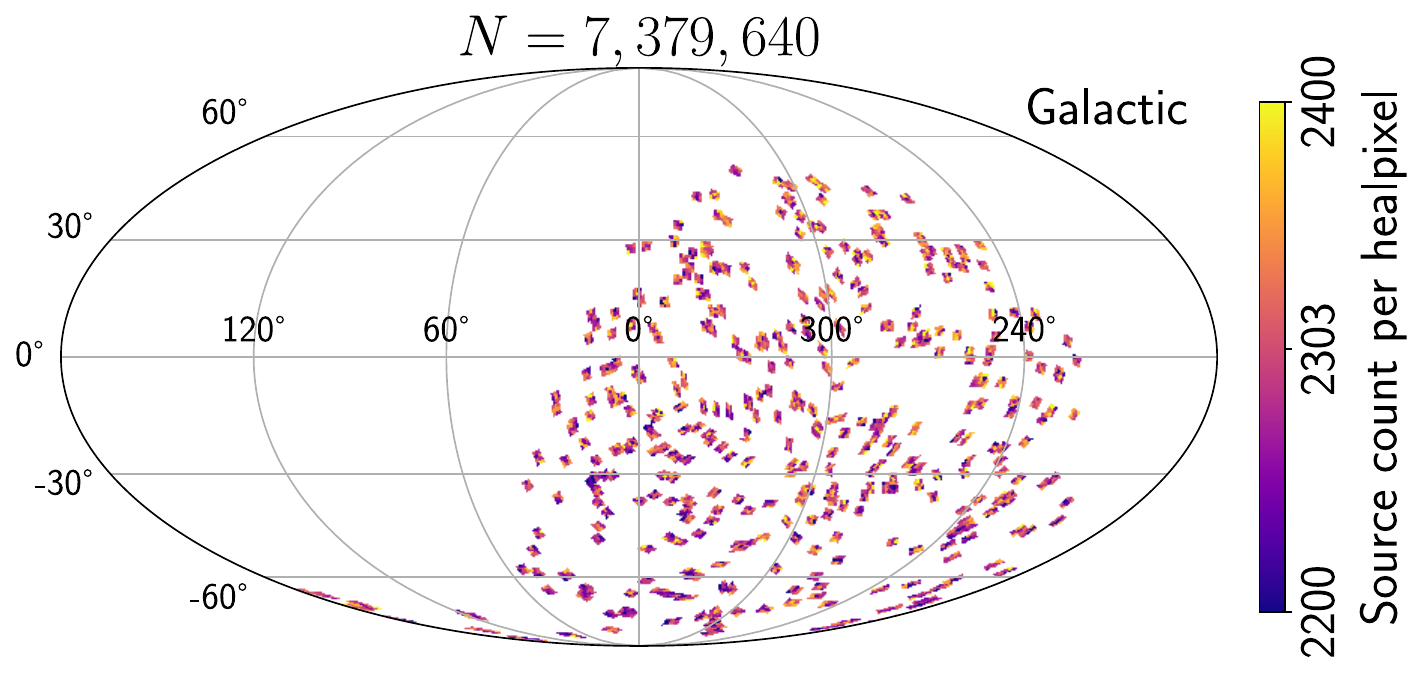}
    \includegraphics[width=\linewidth]{
        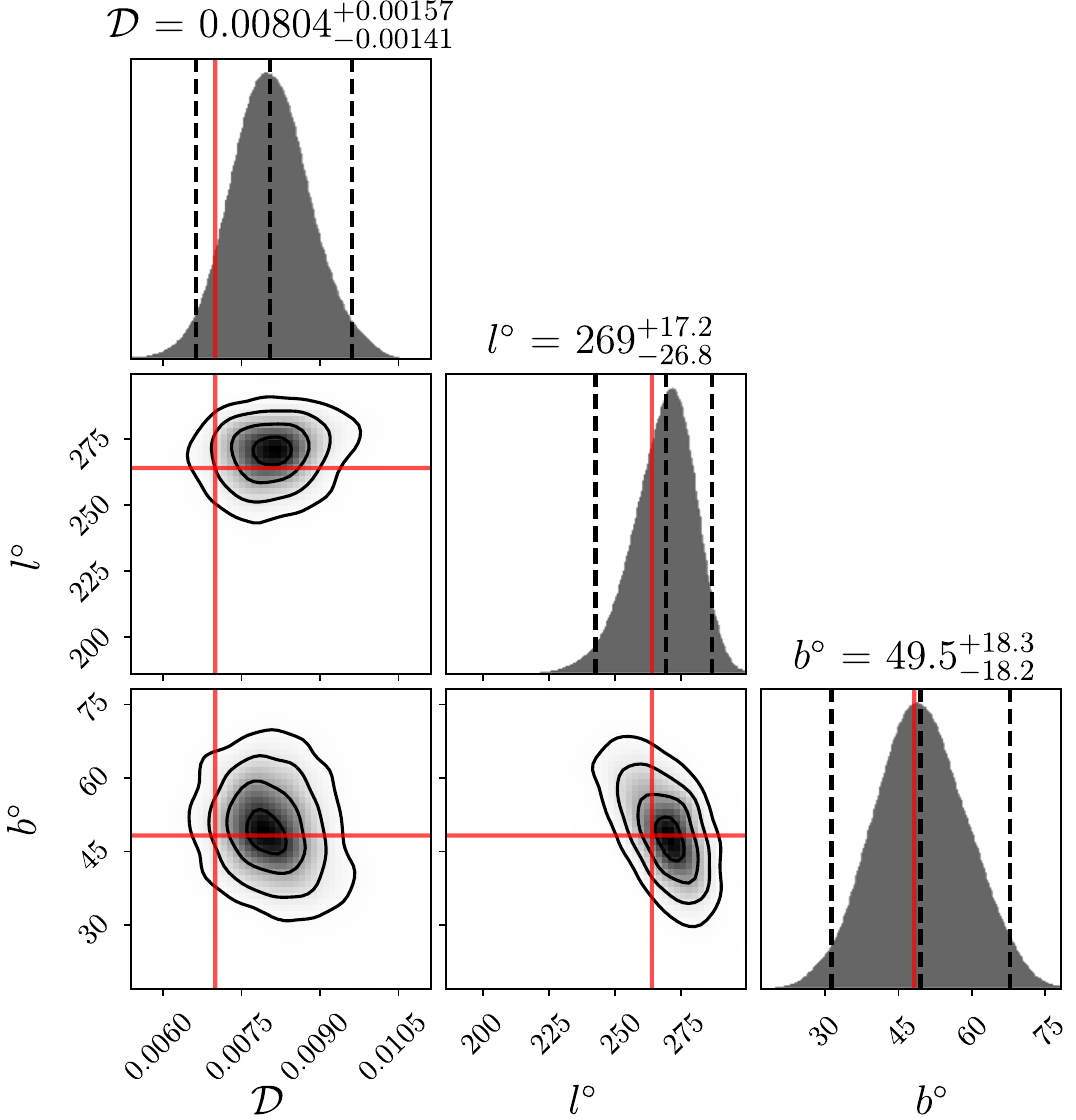
    }
    
    \hfill\includegraphics[width=0.9\linewidth]{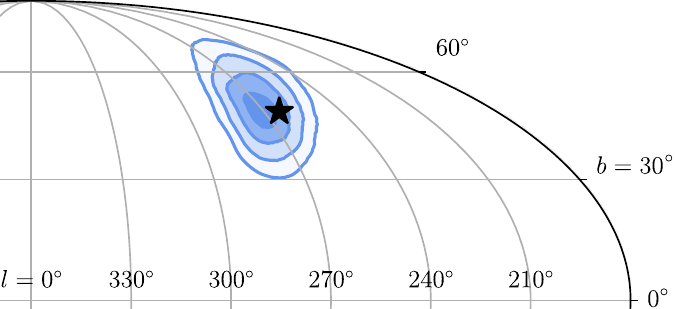}
    \caption{Analysis with a scattered mask.
            \textit{Top:} Visible and masked regions of the map in equatorial
            coordinates.
            \textit{Middle:} Corner plot of the posterior for a dipole fit ($M_1$)
            to the above sample, as in Fig.~\ref{fig:degen_slice}.
            \textit{Bottom:} Sky projection of dipole direction distribution,
            as in Fig.~\ref{fig:degen_slice}.}
    \label{fig:rand_scatter}
\end{figure}

Another possibility to consider is if the visible sky is not contiguous
with the masked region, i.e. the survey has scattered `chunks' of sky
which do not necessarily overlap.
To test this, we take a pair of $(N, r^\circ)$ and corresponding
$D_{\text{KL}}$ from Fig.~\ref{fig:inf_matrix}.
We then create a non-continuous mask by randomly choosing pixels
below a declination of $\delta^\circ = r^\circ - 90^\circ$ (i.e. all
the selected pixels lie within the continuous slice, as defined earlier).
For each selected pixel, we also include all neighbouring pixels.
The set of all these pixels constitutes the visible sky; the remainder is masked.
Lastly, we use a mean source density such that the total number of sources
is approximately equal to $N$.
As an example, we plot the case of $N=7.5$ million, $r^\circ=80$ and $D_{\text{KL}}=5.5$
in Fig.~\ref{fig:rand_scatter}.
391 central pixels (not inclusive of neighbours) were used,
imposing a mean source density of $\bar{\mathcal{N}}\approx 2300$.
This specific number of central pixels was motivated by \citet{wagenveld2024},
in which a dipole measure was performed
on the 391 individual pointings of the
MeerKAT Absorption Line Survey \citep[][; MALS]{meerkat}.
Interestingly, as evident from the corner plot and sky projection,
the dipole parameters can still be robustly recovered despite
the scattered nature of the mask.
Moreover, repeating this analysis 20 times with the same mask parameters
but resampled cell counts, we find the mean KL divergence is
$D_{\text{KL}} = 5.5 \pm 0.1$ (1$\sigma$ uncertainties),
consistent with the KL divergence for the same set of parameters
but a continuous region of sky.
Also, all the Bayes factors are well beyond the `overwhelming' threshold of
$\ln B_{10} = 5$.

This tell us a few things. First, the discontinuous nature of the mask
does not pose an additional hurdle to inference.
Second, the angular extent of the survey (which we call $r^\circ_{\text{sky}}$)
seems to have a more material impact on the information,
as opposed to the fraction of visible sky.
This is because the scattered mask covers about half as much
of the celestial sphere compared to the continuous mask,
whereas $N$ and the angular breadth of the visible sky in declination are fixed.

\subsection{Effect of priors: a cautionary tale}
\label{sub:effect_priors}
We conclude our results with a final remark about the sensitivity of one's
inferences to their choice of prior.
This is a critical feature of Bayesian inference,
and can be seen explicitly in \eqref{eq:bayes-theorem}.
By specifying different prior likelihood functions $\pi(\Theta | M)$,
we can determine the degree to which the final conclusions
are sensitive to the assumptions made before arrival of the data.

Consider the prior choice we made,
as given in Section~\ref{subs:param_optim}.
To explore the different possible dipoles that could explain
the synthetic data,
we sample directions ($l$ and $b$) that are uniform over the sphere.
In other words, we keep the probability per unit area the same,
taking care of how the area element changes with polar angle
in spherical coordinates.
Alternatively, one could construct a 3-parameter model
by defining the dipole
vector in Cartesian coordinates, where $\mathbf{d} = (d_x, d_y, d_z)$.
From this prescription,
it may seem like a logical choice to sample the components of the
dipole with uniform distributions, e.g. $d_x \sim \mathcal{U}(-0.1, 0.1)$.
However,
this choice of prior -- which at first seems like a typical ignorance prior
-- in fact puts strong prior constraints on the dipole amplitude. 

To illustrate this, we performed another set of simulations.
We first constructed an unmasked monopole sample
using $\approx$~1,000,000 sources;
that is, our sample contains no intrinsic dipole or quadrupole,
but a uniform source density over all $l$ and $b$.
We then fit a dipole to the sample with two different approaches.
In the first approach, our choice of prior is the same
as given in Section~\ref{subs:param_optim}.
In the second approach, we take flat priors on all the dipole components
in Cartesian coordinates: $d_x, d_y, d_z \sim \mathcal{U}(-0.1, 0.1)$.
This captures both the dipole direction and magnitude.

\begin{figure}
    \centering
    \includegraphics[width=\linewidth]{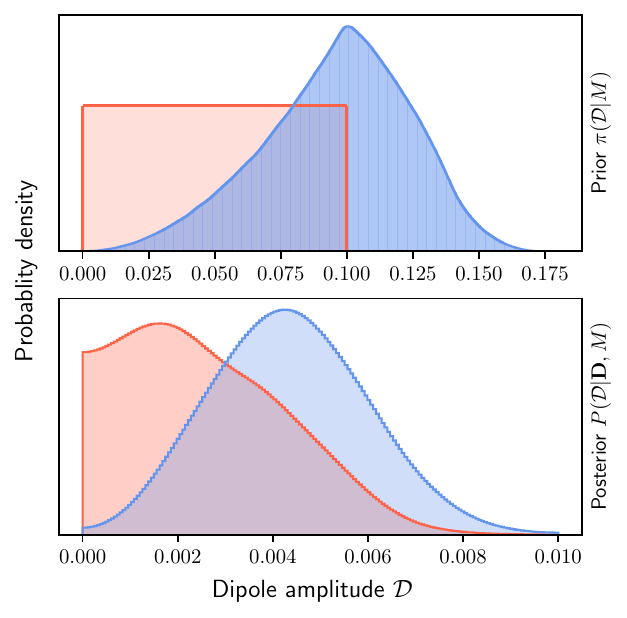}
    \caption{Probability distributions using a flat prior on the
             dipole amplitude (red) and flat priors on the Cartesian
             components of the dipole vector (blue).
             For the sake of visualisation,
             all distributions (except the uniform one)
             have been smoothed through convolution with a Gaussian kernel.
             \textit{Top:} Prior probability distributions
             for $\mathcal{D}$. \textit{Bottom:} Posterior distributions for
             $\mathcal{D}$ after fitting a dipole with either prior
             to a synthetic monopole sample with 1,000,000 sources.}
    \label{fig:choice_of_prior}
\end{figure}

Our results are shown in Fig.~\ref{fig:choice_of_prior}.
The top panel illustrates the prior probability distribution
for both approaches.
The red rectangle reflects the uniform distribution we use
for the dipole amplitude, namely $\mathcal{D} \sim \mathcal{U}(0, 0.1)$.
Meanwhile,
the blue distribution reflects the prior on the dipole amplitude
where one samples the Cartesian components uniformly.
We generated this distribution by randomly sampling
$d_x$, $d_y$ and $d_z$
and then plotting the distribution of
$\mathcal{D} = \sqrt{d_x^2 + d_y^2 + d_z^2}$.
By inspection, sampling in this manner
puts an a priori weighting on amplitudes near 0.1,
while lower amplitudes have near 0 probability density.
Looking at the posterior distributions
for the dipole amplitude (bottom panel),
this a priori weighting translates to small amplitudes
also having a lower probability density for the blue distribution.
Conversely, for the red distribution,
most of the probability mass is concentrated near lower amplitudes.
Recall that the sample we used \textit{only consisted of a monopole}.
We would therefore hope that, after inference,
the zero amplitude has near the largest probability density.
This is the case for our choice of prior,
but not where one samples the Cartesian dipole components uniformly.

This result illustrates the importance of carefully selecting one's prior,
including the importance of considering any information
contained in the prior which might seem absent prima facie.
We stress that this feature is not unique to Bayesian inference.
Bayes's theorem provides a way for explicitly
formulating prior assumptions, and thus for independent studies
to codify their potentially different states of knowledge before the data.
A frequentist-style analysis (e.g. a maximum likelihood estimate)
would still suffer from the same issue if, for example,
the likelihood function involves choosing
the Cartesian dipole components uniformly.
As an example,
\textsc{healpy}'s \texttt{fit\_dipole} function uses a least-squares
estimator of the dipole amplitude,
minimising the sum of the residuals
between the data (the observed pixel counts) and the model
(the sum of a monopole and orthogonal dipoles $d_x$, $d_y$ and $d_z$).
This function was used for instance in \citet{secrest2021,secrest2022}.
We tested this function on the same monopole sample as used above,
finding that $\mathcal{D} \approx 0.0035$.
This is reasonably close to the peak of the blue distribution
in Fig.~\ref{fig:choice_of_prior}.

\begin{figure}
    \centering
    \includegraphics[width=\linewidth]{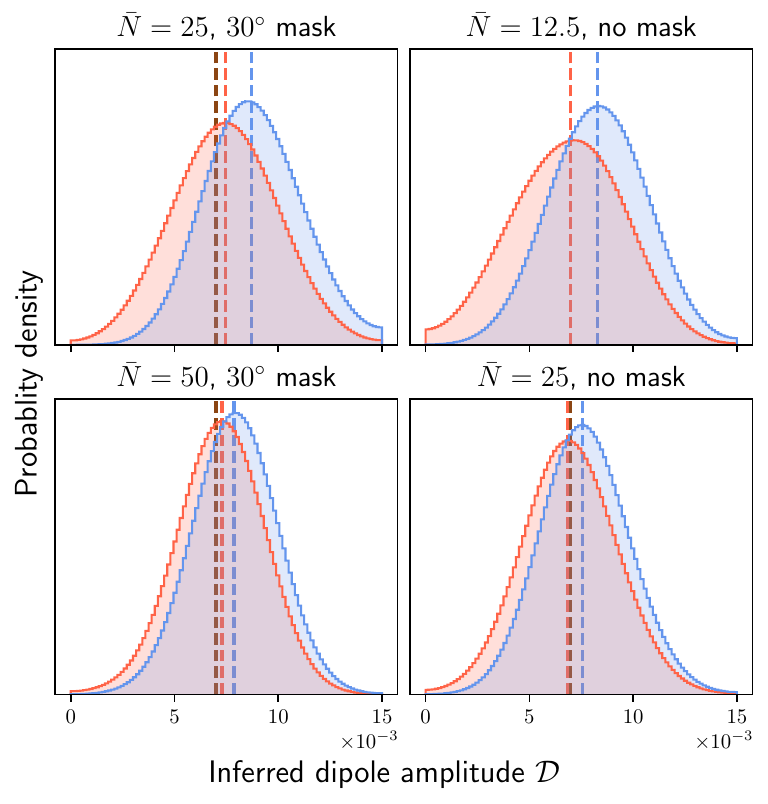}
    \caption{Posterior probability distributions (smoothed with a Gaussian kernel)
            for the dipole amplitude
            using a flat prior on the amplitude (red) and flat priors
            on the Cartesian components of the dipole vector (blue).
            The red and blue dashed lines indicate the medians of each
            distribution. The brown dashed line indicates the true
            dipole amplitude as embedded in the synthetic sample
            ($\mathcal{D} \approx 0.007$). The following describes
            the variations used on the synthetic sample.
            \textit{Left column:} $30^\circ$ galactic plane mask.
            \textit{Right column:} No galactic plane mask.
            \textit{Top row:} $N \approx$ 600,000 sources.
            \textit{Bottom row:} $N \approx$ 1,200,000 sources.}
    \label{fig:choice_of_prior2}
\end{figure}

To probe the issue further,
we repeated our above analysis except with a dipole sample
($\mathcal{D} = 0.007$)
instead of a monopole sample.
After each run, we recorded the median inferred dipole amplitude.
We also constructed four different samples,
running 200 simulations with each: 
\begin{itemize}
    \item $\bar{\mathcal{N}\mathcal{}} = 25$, $|b| \leq 30^\circ$ masked, $N \approx$ 600,000.
    \item $\bar{\mathcal{N}} = 12.5$, no mask, $N \approx$ 600,000.
    \item $\bar{\mathcal{N}} = 50$, $|b| \leq 30^\circ$ masked, $N \approx$ 1,200,000.
    \item $\bar{\mathcal{N}} = 25$, no mask, $N \approx$ 1,200,000.
\end{itemize}
Our results are shown in Fig.~\ref{fig:choice_of_prior2}.
By inspection,
the blue distribution (uniform Cartesian component)
prefers higher values for the dipole amplitude than the red distribution
(uniform over the sphere).
One can also see this by the blue and red dashed lines,
which denote the median of each distribution.
It is also worth noting that,
with more data -- that is, with more sources as in the bottom row
of the figure -- the medians of either distribution converge
to the true dipole amplitude.
This is as we would expect,
since the arrival of more informative data washes out
the initial effect of the choice of prior.
Nonetheless,
in the top row,
the difference between the true amplitude
and the median inferred amplitude for the blue distribution
is non-negligible
($\mathcal{D} = 0.007$ vs. $\mathcal{D} \approx 0.0085$.),
and is a important example of how one needs to be
careful in choosing their prior.

\section{Discussion \& Conclusions}
\label{sec:discussion_conclusions}
\subsection{Main Findings}
The dipole tension represents a growing challenge to the kinematic
interpretation of the CMB, and by extension the cosmological principle.
Because of the significant number of independent studies finding
support for an anomalously large dipole, it is genuinely worth
probing the methodology used to infer this dipole --
especially with an eye to seeing if certain methods suffer
from biases or other pitfalls.
In this work, we applied our Bayesian statistical approach,
as well as the traceless symmetric tensor method,
to samples with significant masks applied and a spectrum 
of different source counts.
Some of these samples included higher order multipoles -- in addition
to a dipole -- up to $\ell = 3$.
To digest our findings,
we give a list of propositions drawn from the totality of the results.

\begin{proposition}
    \label{prop:info-content}
    The information content of the data is chiefly determined by the
    number of sources $N$ and the radius of visible sky $r^\circ_{\text{sky}}$.
    Increasing $r^\circ_{\text{sky}}$ at a fixed $N$ increases the amount
    of information per source.
    At a fixed $r^\circ_{\text{sky}}$, increasing $N$ increases the amount of total
    information, but decreases the amount of information per source.
\end{proposition}

\begin{proposition}
    \label{prop:dipole_robust}
    For the pure dipole sample ($S_1$),
    our estimate of intrinsic dipole parameters is robust to masked skies
    until the data holds insufficient information, in which case the dipole
    and monopole models have near-equal odds. With $\mathcal{D}=0.007$,
    we need $N \approx 500\,000$ sources before there is strong positive support
    for a dipole over the null hypothesis.
\end{proposition}

\begin{proposition}
    \label{prop:small_sky}
    Significant coverage of the celestial sphere is not a requirement
    in recovering accurate dipole parameters.
    We can perform inference on small patches of visible sky covering, for example,
    $12\%$ to $41\%$ of the celestial sphere with source counts in the low
    tens of millions (see Fig.~\ref{fig:inf_matrix}).
    This is possible with singular and non-continuous, sparsely-scattered
    elements of the celestial sphere.
\end{proposition}

\begin{proposition}
    \label{prop:mode_mixing}
    We can recover dipole, quadrupole and octupole parameters accurately
    for samples consisting of combinations of these underlying multipole signals,
    even if those samples have significant masks applied
    ($f_{\text{sky}} = 0.5$ with $g_{\text{mask}}^\circ = 30$).
    That is to say, mode coupling on masked skies is not a concern for our approach.
\end{proposition}

\begin{proposition}
    \label{prop:multipole}
    Our mathematical framework can be generalised to higher order multipoles --
    the only limitation being the computational overhead.
\end{proposition}

Propositions~\ref{prop:info-content} and \ref{prop:small_sky}
imply that a near all-sky survey is not a requirement as far as
accurate dipole estimation is concerned.
In Proposition~\ref{prop:info-content},
$r^\circ_{\text{sky}}$ refers to the angular
breadth of the visible region of sky (see, for example, Fig.~\ref{fig:bmap_slices}).
We use this term since the region of sky need not be continuous, as was shown in
Fig.~\ref{fig:rand_scatter}, in which the discontinuous patches
below $\delta = -10^\circ$ translates to $r^\circ_{\text{sky}} = 80^\circ$.
Now, one can minimise the total number of sources needed by maximising
$r^\circ_{\text{sky}}$,
though in many cases this is not achievable.
This suggests that it is quite desirable for a
prospective survey to focus on an isolated region where its sensitivity is highest.
Ideally, this would safeguard against any systematic effects which are
correlated with, for instance, declination.
Our findings on this point share some similarities with \citet{yoon2015},
in which a frequentist-style analysis revealed that a survey
covering 75\% of the sky ($g^\circ_{\text{mask}} = 15$)
with $N \gtrapprox 30$ million sources would be sufficient for a
$5\sigma$ detection of the cosmic dipole.
While our studies are consistent in finding that the informativeness of
the survey is contingent not only on source count, but sky coverage
and the orientation of the mask with respect to the underlying signal,
we have additionally shown that we can break the degeneracy between the dipole
and higher order multipoles on partial skies
(Proposition~\ref{prop:mode_mixing}).

An interesting consequence of Proposition~\ref{prop:dipole_robust} is that,
given the typical sample source counts
across various radio dipole studies \citep[see e.g.][]{oayda2024},
the information content of the radio samples would be too limited to infer a dipole
but for the amplitudes being $\times 2$ to $\times 3$ as large
as the expectation of $\mathcal{D}_{\text{CMB}} \approx 0.004$.
This also sheds some light on the result of \citet{wagenveld2024},
where the authors considered the dipole in MALS,
which consists of 391 individual pointings with an angular size of about $3.3^\circ$.
This roughly corresponds to the sparse catalogue we constructed in Fig.~\ref{fig:rand_scatter}.
By the 400~$\mu\text{Jy}$ flux cut,
the authors' sample had approximately $300\,000$ sources
below a declination of $\delta \approx 30^\circ$ across these discrete images. 
This is below the flat $500\,000$ source for dipole detection identified in Section~\ref{sub:dipole_sample},
as well as the 6.6 million threshold for $80^\circ$ slices at
$D_{\text{KL}} = 5.5$.
Thus, it is unsurprising that, while their estimate of the dipole amplitude is consistent
with the CMB expectation, it is also consistent with other results
which report an amplitude in excess,
for example \citet{secrest2021}.

We can obtain a rough estimate of the number of sources that would be
needed in MALS to infer a $3\sigma$ tension between the CMB dipole amplitude
if the dipole amplitude is genuinely twice as large.
To do this, we take the list of the 391 pointings from MALS, using these
as the locations of central pixels. We then query all neighbouring pixels.
With $N_\text{side} = 64$, the side length of this patch of sky defined
by the central pixel and neighbours is roughly $3^\circ$, similar
to the $3.3^\circ$ side length of the MALS pointings.
These patches define our visible sky.
We then iterate through synthetic catalogues constructed with this template
at different source counts and with an intrinsic dipole signal
at $2 \times \mathcal{D}_{\text{CMB}} = 0.008$,
where the CMB expectation $\mathcal{D}_{\text{CMB}}$
has been taken from \citet{wagenveld2024}.
At each iteration, we compute the integral over the marginal distribution
\begin{equation}
    P(\mathcal{D} > \mathcal{D}_{\text{CMB}}) = 
        \int_{\mathcal{D}_{\text{CMB}}}^{\infty}
        P(\mathcal{D} | \mathbf{D}, M_1) \, d\mathcal{D},
\end{equation}
which is the probability that the dipole amplitude
is larger than the CMB expectation.
We then convert this into a statistical significance $S$ in units of $\sigma$
using the normal distribution and look at how the significance
changes as a function of $N$.
We find that by $N \approx 1.2$ million, $S \approx (3.0 \pm 0.9) \sigma$
across the iterations.
This only represents a $\approx \times 4$ increases in source count
compared with the real MALS sample at the 400~$\mu\text{Jy}$ flux cut,
and is much less than the other 6.6 million threshold mentioned above.

On a different note,
motivated by Propositions~\ref{prop:mode_mixing} and \ref{prop:multipole},
it will be worth revisiting \citet{secrest2021} and \citet{secrest2022}
in the future with an eye to disentangling the contributions from the dipole mode
and potentially higher order multipoles.
The ecliptic bias constitutes a strong quadrupole signal,
but in light of the comments in \citet{abghari2024} it will be illuminating
to see if there is still a latent quadrupole.
A more robust characterisation of the higher-order structure in that sample
will shed more light on the nature of the dipole tension,
especially on whether or not it can be explained by these
theoretical multipole-induced errors.

\subsection{Future Surveys}
\label{sub:future}
In the future,
what surveys could reach the requisite source counts and visible sky fractions
that we identified?
We outline a number of these below:
\begin{itemize}
    \item The Square Kilometre Array (SKA)\footnote{\url{https://www.skao.int/en}}
        will provide a wealth of information for probes of the cosmic dipole.
        For example,
        SKA is anticipated to generate a substantial HI 21 cm galaxy survey.
        Phase 1 (SKA1) is expected to cover about 5000 deg$^2$,
        containing 5 million galaxies up to a redshift $z \approx 0.5$.
        Later on, Phase 2 will survey $9\times10^8$ galaxies over a 30,000 deg$^2$ area up to $z\approx2$ \citep{camera2015}.
        While Phase 1 will likely have insufficient sample statistics
        for detection of the cosmic dipole (the 5000 deg$^2$ area is approximately
        equivalent to the $r_{\text{sky}}^\circ = 40$ slice we analysed),
        Phase 2 will be more than sufficient.
        That being said, studies of the cosmic dipole in redshift surveys
        need to take into account additional terms due to redshift-space
        distortions and magnitude perturbations \citep{maartens2018}.
        SKA's HI 21cm intensity mapping survey \citep{santos2015}
        will be similarly affected.
        In terms of SKA's radio continuum survey,
        source counts are projected to be of order $10^8$ with realistic
        sky fractions $f_{\text{sky}} \geq 0.5$ \citep{bengaly2019}.
        This will offer excellent source statistics for a dipole measure
        in the style of \citet{ellis1984}.
    \item The Legacy Survey of Space and Time (LSST),
        observed at the Vera C. Rubin Observatory,
        will survey $\approx 10$ million quasars over the southern sky \citep{LSST2019}.
        With quasars alone, this sets within the threshold of 6.6 million sources
        for $\delta \leq 80^\circ$ we identified.
    \item In terms of current infrastructure,
        the Evolutionary Map of the Universe (EMU) --
        which will use the Australian SKA Pathfinder (ASKAP) --
        is expected to survey about 40 million radio galaxies
        in the southern sky \citep{emu, emu-casda}.
        This is again more than sufficient for a detection of the dipole,
        though the recently-released EMU pilot survey is insufficient
        with about 220,000 sources in a 270~deg$^2$ area \citep{emu-pilot}.
    \item The \textit{Euclid} satellite's Euclid Wide 
        Survey (EWS) will cover a 14,500~deg$^2$ ($f_{\text{sky}} \approx 0.35$)
        area of the celestial sphere, and it is anticipated that
        approximately 40 million AGN will be detected
        in at least one of \textit{Euclid}'s photometric
        bands -- although practically, the number
        of AGN which will actually be selected in the survey
        by simple colour magnitude cuts will be less than
        this \citep{euclid2024}. This lower number is expected to be in
        the vicinity of 5 million AGN. The actual footprint of the EWS
        is non-continuous and not strictly analogous to any mask we have
        used in this work. The closest analogue are the samples with
        $g_{\text{mask}}^\circ=40$, in which case our $N \approx 500\,000$
        threshold for dipole inference is relevant.
        On this basis, Euclid will offer more than sufficient
        sample statistics to probe the cosmic dipole.
\end{itemize}

\section*{Acknowledgements}
This work made use of the \textsc{python} packages
\textsc{dynesty} \citep{skilling2004, skilling2006, dynesty-v2.1.2}, \textsc{healpy} \citep{Gorski2005,Zonca2019},
\textsc{numpy} \citep{harris2020},
\textsc{matplotlib} \citep{hunter2007},
\textsc{scipy} \citep{scipy2020} and \textsc{astropy} \citep{astropy2022}.
We uncovered the notes of \citet{pirani1965}, which contain the
relation at \eqref{eq:general_traceless_tensor},
thanks to a Mathematics Stack Exchange thread.%
\footnote{\url{https://math.stackexchange.com/questions/4925863/finding-general-expression-for-symmetric-trace-free-tensors-stfs}}
The lecture notes of \citet{guth2012_l8, guth2012_l9} are publicly available;
we accessed these with a Google search.
We extend our gratitude to Sebastian von Hausegger
for helpful discussion surrounding the unit vector approach
to multipoles and for directing us to references on its history.
We also thank Tara Murphy for discussions and comments
surrounding this work
and we thank the anonymous referee for their insightful feedback
that improved this paper.
VM is supported by the University of Sydney's Physics Foundation Scholarship.
OTO is supported by the University of Sydney Postgraduate Award.

\section*{Data Availability}
The data used in this study will be made available with
a reasonable request to the authors.



\bibliographystyle{mnras}
\bibliography{main} 

\begin{thebibliography}{}
\makeatletter
\relax
\def\mn@urlcharsother{\let\do\@makeother \do\$\do\&\do\#\do\^\do\_\do\%\do\~}
\def\mn@doi{\begingroup\mn@urlcharsother \@ifnextchar [ {\mn@doi@} {\mn@doi@[]}}
\def\mn@doi@[#1]#2{\def\@tempa{#1}\ifx\@tempa\@empty \href {http://dx.doi.org/#2} {doi:#2}\else \href {http://dx.doi.org/#2} {#1}\fi \endgroup}
\def\mn@eprint#1#2{\mn@eprint@#1:#2::\@nil}
\def\mn@eprint@arXiv#1{\href {http://arxiv.org/abs/#1} {{\tt arXiv:#1}}}
\def\mn@eprint@dblp#1{\href {http://dblp.uni-trier.de/rec/bibtex/#1.xml} {dblp:#1}}
\def\mn@eprint@#1:#2:#3:#4\@nil{\def\@tempa {#1}\def\@tempb {#2}\def\@tempc {#3}\ifx \@tempc \@empty \let \@tempc \@tempb \let \@tempb \@tempa \fi \ifx \@tempb \@empty \def\@tempb {arXiv}\fi \@ifundefined {mn@eprint@\@tempb}{\@tempb:\@tempc}{\expandafter \expandafter \csname mn@eprint@\@tempb\endcsname \expandafter{\@tempc}}}

\bibitem[\protect\citeauthoryear{{Abghari}, {Bunn}, {Hergt}, {Li}, {Scott}, {Sullivan}  \& {Wei}}{{Abghari} et~al.}{2024}]{abghari2024}
{Abghari} A.,  {Bunn} E.~F.,  {Hergt} L.~T.,  {Li} B.,  {Scott} D.,  {Sullivan} R.~M.,   {Wei} D.,  2024, \mn@doi [arXiv e-prints] {10.48550/arXiv.2405.09762}, \href {https://ui.adsabs.harvard.edu/abs/2024arXiv240509762A} {p. arXiv:2405.09762}

\bibitem[\protect\citeauthoryear{{Astropy Collaboration} et~al.,}{{Astropy Collaboration} et~al.}{2022}]{astropy2022}
{Astropy Collaboration} et~al., 2022, \mn@doi [apj] {10.3847/1538-4357/ac7c74}, \href {https://ui.adsabs.harvard.edu/abs/2022ApJ...935..167A} {935, 167}

\bibitem[\protect\citeauthoryear{{Baleisis}, {Lahav}, {Loan}  \& {Wall}}{{Baleisis} et~al.}{1998}]{baleisis1998}
{Baleisis} A.,  {Lahav} O.,  {Loan} A.~J.,   {Wall} J.~V.,  1998, \mn@doi [\mnras] {10.1046/j.1365-8711.1998.01536.x}, \href {https://ui.adsabs.harvard.edu/abs/1998MNRAS.297..545B} {297, 545}

\bibitem[\protect\citeauthoryear{Bengaly, Maartens  \& Santos}{Bengaly et~al.}{2018}]{bengaly2018}
Bengaly C.~A.,  Maartens R.,   Santos M.~G.,  2018, \mn@doi [\jcap] {10.1088/1475-7516/2018/04/031}, 2018, 031

\bibitem[\protect\citeauthoryear{{Bengaly}, {Siewert}, {Schwarz}  \& {Maartens}}{{Bengaly} et~al.}{2019}]{bengaly2019}
{Bengaly} C.~A.~P.,  {Siewert} T.~M.,  {Schwarz} D.~J.,   {Maartens} R.,  2019, \mn@doi [\mnras] {10.1093/mnras/stz832}, \href {https://ui.adsabs.harvard.edu/abs/2019MNRAS.486.1350B} {486, 1350}

\bibitem[\protect\citeauthoryear{{Blake} \& {Wall}}{{Blake} \& {Wall}}{2002}]{blake2002}
{Blake} C.,  {Wall} J.,  2002, \mn@doi [\nat] {10.1038/416150a}, \href {https://ui.adsabs.harvard.edu/abs/2002Natur.416..150B} {416, 150}

\bibitem[\protect\citeauthoryear{{Buchner}}{{Buchner}}{2016}]{buchner2016}
{Buchner} J.,  2016, \mn@doi [Statistics and Computing] {10.1007/s11222-014-9512-y}, \href {https://ui.adsabs.harvard.edu/abs/2016S&C....26..383B} {26, 383}

\bibitem[\protect\citeauthoryear{{Buchner}}{{Buchner}}{2019}]{buchner2019}
{Buchner} J.,  2019, \mn@doi [\pasp] {10.1088/1538-3873/aae7fc}, \href {https://ui.adsabs.harvard.edu/abs/2019PASP..131j8005B} {131, 108005}

\bibitem[\protect\citeauthoryear{{Buchner}}{{Buchner}}{2021}]{ultranest}
{Buchner} J.,  2021, \mn@doi [The Journal of Open Source Software] {10.21105/joss.03001}, \href {https://ui.adsabs.harvard.edu/abs/2021JOSS....6.3001B} {6, 3001}

\bibitem[\protect\citeauthoryear{{Buchner}}{{Buchner}}{2023}]{buchner2023}
{Buchner} J.,  2023, \mn@doi [Statistics Surveys] {10.1214/23-SS144}, \href {https://ui.adsabs.harvard.edu/abs/2023StSur..17..169B} {17, 169}

\bibitem[\protect\citeauthoryear{{Camera}, {Santos}  \& {Maartens}}{{Camera} et~al.}{2015}]{camera2015}
{Camera} S.,  {Santos} M.~G.,   {Maartens} R.,  2015, \mn@doi [\mnras] {10.1093/mnras/stv040}, \href {https://ui.adsabs.harvard.edu/abs/2015MNRAS.448.1035C} {448, 1035}

\bibitem[\protect\citeauthoryear{{Coles} \& {Lucchin}}{{Coles} \& {Lucchin}}{2002}]{coles2002}
{Coles} P.,  {Lucchin} F.,  2002, {Cosmology: The Origin and Evolution of Cosmic Structure, Second Edition}

\bibitem[\protect\citeauthoryear{{Colin}, {Mohayaee}, {Rameez}  \& {Sarkar}}{{Colin} et~al.}{2017}]{colin2017}
{Colin} J.,  {Mohayaee} R.,  {Rameez} M.,   {Sarkar} S.,  2017, \mn@doi [\mnras] {10.1093/mnras/stx1631}, \href {https://ui.adsabs.harvard.edu/abs/2017MNRAS.471.1045C} {471, 1045}

\bibitem[\protect\citeauthoryear{{Condon}, {Cotton}, {Greisen}, {Yin}, {Perley}, {Taylor}  \& {Broderick}}{{Condon} et~al.}{1998}]{nvss-survey}
{Condon} J.~J.,  {Cotton} W.~D.,  {Greisen} E.~W.,  {Yin} Q.~F.,  {Perley} R.~A.,  {Taylor} G.~B.,   {Broderick} J.~J.,  1998, \mn@doi [\aj] {10.1086/300337}, \href {https://ui.adsabs.harvard.edu/abs/1998AJ....115.1693C} {115, 1693}

\bibitem[\protect\citeauthoryear{{Copi}, {Huterer}  \& {Starkman}}{{Copi} et~al.}{2004}]{copi2024}
{Copi} C.~J.,  {Huterer} D.,   {Starkman} G.~D.,  2004, \mn@doi [\prd] {10.1103/PhysRevD.70.043515}, \href {https://ui.adsabs.harvard.edu/abs/2004PhRvD..70d3515C} {70, 043515}

\bibitem[\protect\citeauthoryear{Dalang \& Bonvin}{Dalang \& Bonvin}{2022}]{dalang2022}
Dalang C.,  Bonvin C.,  2022, \mn@doi [Monthly Notices of the Royal Astronomical Society] {10.1093/mnras/stac726}, 512, 3895

\bibitem[\protect\citeauthoryear{{Ellis} \& {Baldwin}}{{Ellis} \& {Baldwin}}{1984}]{ellis1984}
{Ellis} G.~F.~R.,  {Baldwin} J.~E.,  1984, \mn@doi [\mnras] {10.1093/mnras/206.2.377}, \href {https://ui.adsabs.harvard.edu/abs/1984MNRAS.206..377E} {206, 377}

\bibitem[\protect\citeauthoryear{{Euclid Collaboration} et~al.,}{{Euclid Collaboration} et~al.}{2024}]{euclid2024}
{Euclid Collaboration} et~al., 2024, \mn@doi [arXiv e-prints] {10.48550/arXiv.2405.18126}, \href {https://ui.adsabs.harvard.edu/abs/2024arXiv240518126E} {p. arXiv:2405.18126}

\bibitem[\protect\citeauthoryear{{Gibelyou} \& {Huterer}}{{Gibelyou} \& {Huterer}}{2012}]{gibelyou2012}
{Gibelyou} C.,  {Huterer} D.,  2012, \mn@doi [\mnras] {10.1111/j.1365-2966.2012.22032.x}, \href {https://ui.adsabs.harvard.edu/abs/2012MNRAS.427.1994G} {427, 1994}

\bibitem[\protect\citeauthoryear{{G{\'o}rski}, {Hivon}, {Banday}, {Wandelt}, {Hansen}, {Reinecke}  \& {Bartelmann}}{{G{\'o}rski} et~al.}{2005}]{Gorski2005}
{G{\'o}rski} K.~M.,  {Hivon} E.,  {Banday} A.~J.,  {Wandelt} B.~D.,  {Hansen} F.~K.,  {Reinecke} M.,   {Bartelmann} M.,  2005, \mn@doi [\apj] {10.1086/427976}, \href {http://adsabs.harvard.edu/abs/2005ApJ...622..759G} {622, 759}

\bibitem[\protect\citeauthoryear{{Guandalin}, {Piat}, {Clarkson}  \& {Maartens}}{{Guandalin} et~al.}{2023}]{gaundalin2023}
{Guandalin} C.,  {Piat} J.,  {Clarkson} C.,   {Maartens} R.,  2023, \mn@doi [\apj] {10.3847/1538-4357/acdf46}, \href {https://ui.adsabs.harvard.edu/abs/2023ApJ...953..144G} {953, 144}

\bibitem[\protect\citeauthoryear{{Gupta} et~al.,}{{Gupta} et~al.}{2016}]{meerkat}
{Gupta} N.,  et~al., 2016, in MeerKAT Science: On the Pathway to the SKA. p.~14 (\mn@eprint {arXiv} {1708.07371}), \mn@doi{10.22323/1.277.0014}

\bibitem[\protect\citeauthoryear{{Guth}}{{Guth}}{2012a}]{guth2012_l8}
{Guth} A.,  2012a, LECTURE NOTES 8: THE TRACELESS SYMMETRIC TENSOR EXPANSION AND STANDARD SPHERICAL HARMONICS, \url {https://ocw.mit.edu/courses/8-07-electromagnetism-ii-fall-2012/6a67040c6a90f70c8164cdea63473a30_MIT8_07F12_ln8.pdf}

\bibitem[\protect\citeauthoryear{{Guth}}{{Guth}}{2012b}]{guth2012_l9}
{Guth} A.,  2012b, LECTURE NOTES 9: TRACELESS SYMMETRIC TENSOR APPROACH TO LEGENDRE POLYNOMIALS AND SPHERICAL HARMONICS, \url {https://ocw.mit.edu/courses/8-07-electromagnetism-ii-fall-2012/0ccc978da12536a8d95333ec7726c555_MIT8_07F12_ln9.pdf}

\bibitem[\protect\citeauthoryear{{Handley} \& {Lemos}}{{Handley} \& {Lemos}}{2019a}]{handley2019a}
{Handley} W.,  {Lemos} P.,  2019a, \mn@doi [\prd] {10.1103/PhysRevD.100.023512}, \href {https://ui.adsabs.harvard.edu/abs/2019PhRvD.100b3512H} {100, 023512}

\bibitem[\protect\citeauthoryear{{Handley} \& {Lemos}}{{Handley} \& {Lemos}}{2019b}]{handley2019b}
{Handley} W.,  {Lemos} P.,  2019b, \mn@doi [\prd] {10.1103/PhysRevD.100.043504}, \href {https://ui.adsabs.harvard.edu/abs/2019PhRvD.100d3504H} {100, 043504}

\bibitem[\protect\citeauthoryear{Harris et~al.,}{Harris et~al.}{2020}]{harris2020}
Harris C.~R.,  et~al., 2020, \mn@doi [Nature] {10.1038/s41586-020-2649-2}, 585, 357

\bibitem[\protect\citeauthoryear{{Harrison}}{{Harrison}}{2000}]{harrison2000}
{Harrison} E.~R.,  2000, {Cosmology. The science of the universe.}.
Cambridge University Press

\bibitem[\protect\citeauthoryear{{Hopkins}, {Norris}, {Vernstrom}, {Kapinska}  \& {Marvil}}{{Hopkins} et~al.}{2022}]{emu-casda}
{Hopkins} A.,  {Norris} R.,  {Vernstrom} T.,  {Kapinska} A.,   {Marvil} J.,  2022, ASKAP Data Products for Project AS201 (EMU): images and visibilities. v1., \url {ttp://hdl.handle.net/102.100.100/479788?index=1}

\bibitem[\protect\citeauthoryear{{Hu}, {Jia}, {Hu}  \& {Wang}}{{Hu} et~al.}{2024}]{hu2024}
{Hu} J.~P.,  {Jia} X.~D.,  {Hu} J.,   {Wang} F.~Y.,  2024, \mn@doi [arXiv e-prints] {10.48550/arXiv.2410.06450}, \href {https://ui.adsabs.harvard.edu/abs/2024arXiv241006450H} {p. arXiv:2410.06450}

\bibitem[\protect\citeauthoryear{Hunter}{Hunter}{2007}]{hunter2007}
Hunter J.~D.,  2007, \mn@doi [Computing in Science \& Engineering] {10.1109/MCSE.2007.55}, 9, 90

\bibitem[\protect\citeauthoryear{Ivezi\`{c} et~al.,}{Ivezi\`{c} et~al.}{2019}]{LSST2019}
Ivezi\`{c} {\v{Z}}.,  et~al., 2019, \mn@doi [\apj] {10.3847/1538-4357/ab042c}, 873, 111

\bibitem[\protect\citeauthoryear{Kass \& Raftery}{Kass \& Raftery}{1995}]{kass1995}
Kass R.~E.,  Raftery A.~E.,  1995, Journal of the American Statistical Association, 90, 773

\bibitem[\protect\citeauthoryear{Koposov et~al.,}{Koposov et~al.}{2023b}]{dynesty-v2.1.2}
Koposov S.,  et~al., 2023b, joshspeagle/dynesty: v2.1.2, \mn@doi{10.5281/zenodo.7995596}, \url {https://doi.org/10.5281/zenodo.7995596}

\bibitem[\protect\citeauthoryear{Koposov et~al.,}{Koposov et~al.}{2023a}]{dynesty-v2.1.3}
Koposov S.,  et~al., 2023a, joshspeagle/dynesty: v2.1.3, \mn@doi{10.5281/zenodo.8408702}, \url {https://doi.org/10.5281/zenodo.8408702}

\bibitem[\protect\citeauthoryear{{Kumar Aluri} et~al.,}{{Kumar Aluri} et~al.}{2023}]{aluri2023}
{Kumar Aluri} P.,  et~al., 2023, \mn@doi [Classical and Quantum Gravity] {10.1088/1361-6382/acbefc}, \href {https://ui.adsabs.harvard.edu/abs/2023CQGra..40i4001K} {40, 094001}

\bibitem[\protect\citeauthoryear{{Land} et~al.,}{{Land} et~al.}{2008}]{land2008}
{Land} K.,  et~al., 2008, \mn@doi [\mnras] {10.1111/j.1365-2966.2008.13490.x}, \href {https://ui.adsabs.harvard.edu/abs/2008MNRAS.388.1686L} {388, 1686}

\bibitem[\protect\citeauthoryear{{Maartens}, {Clarkson}  \& {Chen}}{{Maartens} et~al.}{2018}]{maartens2018}
{Maartens} R.,  {Clarkson} C.,   {Chen} S.,  2018, \mn@doi [\jcap] {10.1088/1475-7516/2018/01/013}, \href {https://ui.adsabs.harvard.edu/abs/2018JCAP...01..013M} {2018, 013}

\bibitem[\protect\citeauthoryear{{Marocco} et~al.,}{{Marocco} et~al.}{2021}]{marocco2021}
{Marocco} F.,  et~al., 2021, \mn@doi [\apjs] {10.3847/1538-4365/abd805}, \href {https://ui.adsabs.harvard.edu/abs/2021ApJS..253....8M} {253, 8}

\bibitem[\protect\citeauthoryear{{McConnell} et~al.,}{{McConnell} et~al.}{2020}]{racs-original}
{McConnell} D.,  et~al., 2020, \mn@doi [\pasa] {10.1017/pasa.2020.41}, \href {https://ui.adsabs.harvard.edu/abs/2020PASA...37...48M} {37, e048}

\bibitem[\protect\citeauthoryear{{Mittal}, {Oayda}  \& {Lewis}}{{Mittal} et~al.}{2024}]{mittal2024}
{Mittal} V.,  {Oayda} O.~T.,   {Lewis} G.~F.,  2024, \mn@doi [\mnras] {10.1093/mnras/stad3706}, \href {https://ui.adsabs.harvard.edu/abs/2024MNRAS.527.8497M} {527, 8497}

\bibitem[\protect\citeauthoryear{{Mittal}, {Oayda}  \& {Lewis}}{{Mittal} et~al.}{prep}]{mittalPREP}
{Mittal} V.,  {Oayda} O.~T.,   {Lewis} G.~F.,  in prep.

\bibitem[\protect\citeauthoryear{{Norris} et~al.,}{{Norris} et~al.}{2011}]{emu}
{Norris} R.~P.,  et~al., 2011, \mn@doi [\pasa] {10.1071/AS11021}, \href {https://ui.adsabs.harvard.edu/abs/2011PASA...28..215N} {28, 215}

\bibitem[\protect\citeauthoryear{{Norris} et~al.,}{{Norris} et~al.}{2021}]{emu-pilot}
{Norris} R.~P.,  et~al., 2021, \mn@doi [\pasa] {10.1017/pasa.2021.42}, \href {https://ui.adsabs.harvard.edu/abs/2021PASA...38...46N} {38, e046}

\bibitem[\protect\citeauthoryear{{Oayda} \& {Lewis}}{{Oayda} \& {Lewis}}{2023}]{oayda2023}
{Oayda} O.~T.,  {Lewis} G.~F.,  2023, \mn@doi [\mnras] {10.1093/mnras/stad1454}, \href {https://ui.adsabs.harvard.edu/abs/2023MNRAS.523..667O} {523, 667}

\bibitem[\protect\citeauthoryear{{Oayda}, {Mittal}, {Lewis}  \& {Murphy}}{{Oayda} et~al.}{2024}]{oayda2024}
{Oayda} O.~T.,  {Mittal} V.,  {Lewis} G.~F.,   {Murphy} T.,  2024, \mn@doi [\mnras] {10.1093/mnras/stae1399}, \href {https://ui.adsabs.harvard.edu/abs/2024MNRAS.531.4545O} {531, 4545}

\bibitem[\protect\citeauthoryear{Peebles}{Peebles}{2022}]{Peebles_2022}
Peebles P.,  2022, \mn@doi [Annals of Physics] {10.1016/j.aop.2022.169159}, 447, 169159

\bibitem[\protect\citeauthoryear{{Peebles} \& {Wilkinson}}{{Peebles} \& {Wilkinson}}{1968}]{peebles1968}
{Peebles} P.~J.,  {Wilkinson} D.~T.,  1968, \mn@doi [Physical Review] {10.1103/PhysRev.174.2168}, \href {https://ui.adsabs.harvard.edu/abs/1968PhRv..174.2168P} {174, 2168}

\bibitem[\protect\citeauthoryear{{Pirani}}{{Pirani}}{1965}]{pirani1965}
{Pirani} F.~A.~E.,  1965, in , Vol.~1, Lectures on General Relativity.
Prentice-Hall, pp 249--373

\bibitem[\protect\citeauthoryear{{Planck Collaboration} et~al.,}{{Planck Collaboration} et~al.}{2020}]{planck2020}
{Planck Collaboration} et~al., 2020, \mn@doi [\aap] {10.1051/0004-6361/201833880}, \href {https://ui.adsabs.harvard.edu/abs/2020A&A...641A...1P} {641, A1}

\bibitem[\protect\citeauthoryear{Rubart \& Schwarz}{Rubart \& Schwarz}{2013}]{rubart2013}
Rubart M.,  Schwarz D.~J.,  2013, \mn@doi [\aap] {10.1051/0004-6361/201321215}, 555, A117

\bibitem[\protect\citeauthoryear{{Santos} et~al.,}{{Santos} et~al.}{2015}]{santos2015}
{Santos} M.,  et~al., 2015, in Advancing Astrophysics with the Square Kilometre Array (AASKA14). p.~19 (\mn@eprint {arXiv} {1501.03989}), \mn@doi{10.22323/1.215.0019}

\bibitem[\protect\citeauthoryear{{Schwarz}, {Starkman}, {Huterer}  \& {Copi}}{{Schwarz} et~al.}{2004}]{schwarz2004}
{Schwarz} D.~J.,  {Starkman} G.~D.,  {Huterer} D.,   {Copi} C.~J.,  2004, \mn@doi [\prl] {10.1103/PhysRevLett.93.221301}, \href {https://ui.adsabs.harvard.edu/abs/2004PhRvL..93v1301S} {93, 221301}

\bibitem[\protect\citeauthoryear{{Secrest}, {von Hausegger}, {Rameez}, {Mohayaee}, {Sarkar}  \& {Colin}}{{Secrest} et~al.}{2021}]{secrest2021}
{Secrest} N.~J.,  {von Hausegger} S.,  {Rameez} M.,  {Mohayaee} R.,  {Sarkar} S.,   {Colin} J.,  2021, \mn@doi [\apjl] {10.3847/2041-8213/abdd40}, \href {https://ui.adsabs.harvard.edu/abs/2021ApJ...908L..51S} {908, L51}

\bibitem[\protect\citeauthoryear{{Secrest}, {von Hausegger}, {Rameez}, {Mohayaee}  \& {Sarkar}}{{Secrest} et~al.}{2022}]{secrest2022}
{Secrest} N.~J.,  {von Hausegger} S.,  {Rameez} M.,  {Mohayaee} R.,   {Sarkar} S.,  2022, \mn@doi [\apjl] {10.3847/2041-8213/ac88c0}, \href {https://ui.adsabs.harvard.edu/abs/2022ApJ...937L..31S} {937, L31}

\bibitem[\protect\citeauthoryear{Siewert, Schmidt-Rubart  \& Schwarz}{Siewert et~al.}{2021}]{siewert2021}
Siewert T.~M.,  Schmidt-Rubart M.,   Schwarz D.~J.,  2021, \mn@doi [\aap] {10.1051/0004-6361/202039840}, 653, A9

\bibitem[\protect\citeauthoryear{{Singal}}{{Singal}}{2023}]{singal2023}
{Singal} A.~K.,  2023, \mn@doi [\mnras] {10.1093/mnras/stad2161}, \href {https://ui.adsabs.harvard.edu/abs/2023MNRAS.524.3636S} {524, 3636}

\bibitem[\protect\citeauthoryear{{Skilling}}{{Skilling}}{2004}]{skilling2004}
{Skilling} J.,  2004, in {Fischer} R.,  {Preuss} R.,   {Toussaint} U.~V.,  eds,  American Institute of Physics Conference Series Vol. 735, Bayesian Inference and Maximum Entropy Methods in Science and Engineering: 24th International Workshop on Bayesian Inference and Maximum Entropy Methods in Science and Engineering. pp 395--405, \mn@doi{10.1063/1.1835238}

\bibitem[\protect\citeauthoryear{{Skilling}}{{Skilling}}{2006}]{skilling2006}
{Skilling} J.,  2006, \mn@doi [Bayesian Analysis] {10.1214/06-BA127}, 1, 833

\bibitem[\protect\citeauthoryear{Virtanen et~al.,}{Virtanen et~al.}{2020}]{scipy2020}
Virtanen P.,  et~al., 2020, \mn@doi [Nature Methods] {10.1038/s41592-019-0686-2}, \href {https://rdcu.be/b08Wh} {17, 261}

\bibitem[\protect\citeauthoryear{{Wagenveld}, {Kl\"ockner}  \& {Schwarz}}{{Wagenveld} et~al.}{2023}]{wagenveld2023}
{Wagenveld} J.~D.,  {Kl\"ockner} H.-R.,   {Schwarz} D.~J.,  2023, \mn@doi [A\&A] {10.1051/0004-6361/202346210}, 675, A72

\bibitem[\protect\citeauthoryear{{Wagenveld} et~al.,}{{Wagenveld} et~al.}{2024}]{wagenveld2024}
{Wagenveld} J.~D.,  et~al., 2024, \mn@doi [arXiv e-prints] {10.48550/arXiv.2408.16619}, \href {https://ui.adsabs.harvard.edu/abs/2024arXiv240816619W} {p. arXiv:2408.16619}

\bibitem[\protect\citeauthoryear{{Weeks}}{{Weeks}}{2004}]{weeks2004}
{Weeks} J.~R.,  2004, \mn@doi [arXiv e-prints] {10.48550/arXiv.astro-ph/0412231}, \href {https://ui.adsabs.harvard.edu/abs/2004astro.ph.12231W} {pp astro--ph/0412231}

\bibitem[\protect\citeauthoryear{{Yoon} \& {Huterer}}{{Yoon} \& {Huterer}}{2015}]{yoon2015}
{Yoon} M.,  {Huterer} D.,  2015, \mn@doi [\apjl] {10.1088/2041-8205/813/1/L18}, \href {https://ui.adsabs.harvard.edu/abs/2015ApJ...813L..18Y} {813, L18}

\bibitem[\protect\citeauthoryear{Zonca, Singer, Lenz, Reinecke, Rosset, Hivon  \& Gorski}{Zonca et~al.}{2019}]{Zonca2019}
Zonca A.,  Singer L.,  Lenz D.,  Reinecke M.,  Rosset C.,  Hivon E.,   Gorski K.,  2019, \mn@doi [Journal of Open Source Software] {10.21105/joss.01298}, 4, 1298

\bibitem[\protect\citeauthoryear{{von Hausegger}}{{von Hausegger}}{2024}]{vonHausegger2024}
{von Hausegger} S.,  2024, \mn@doi [\mnras] {10.1093/mnrasl/slae092}, \href {https://ui.adsabs.harvard.edu/abs/2024MNRAS.535L..49V} {535, L49}

\makeatother
\end{thebibliography}




\appendix
\section{Additional material}
\begin{figure}
    \centering
    \includegraphics[width=\linewidth]{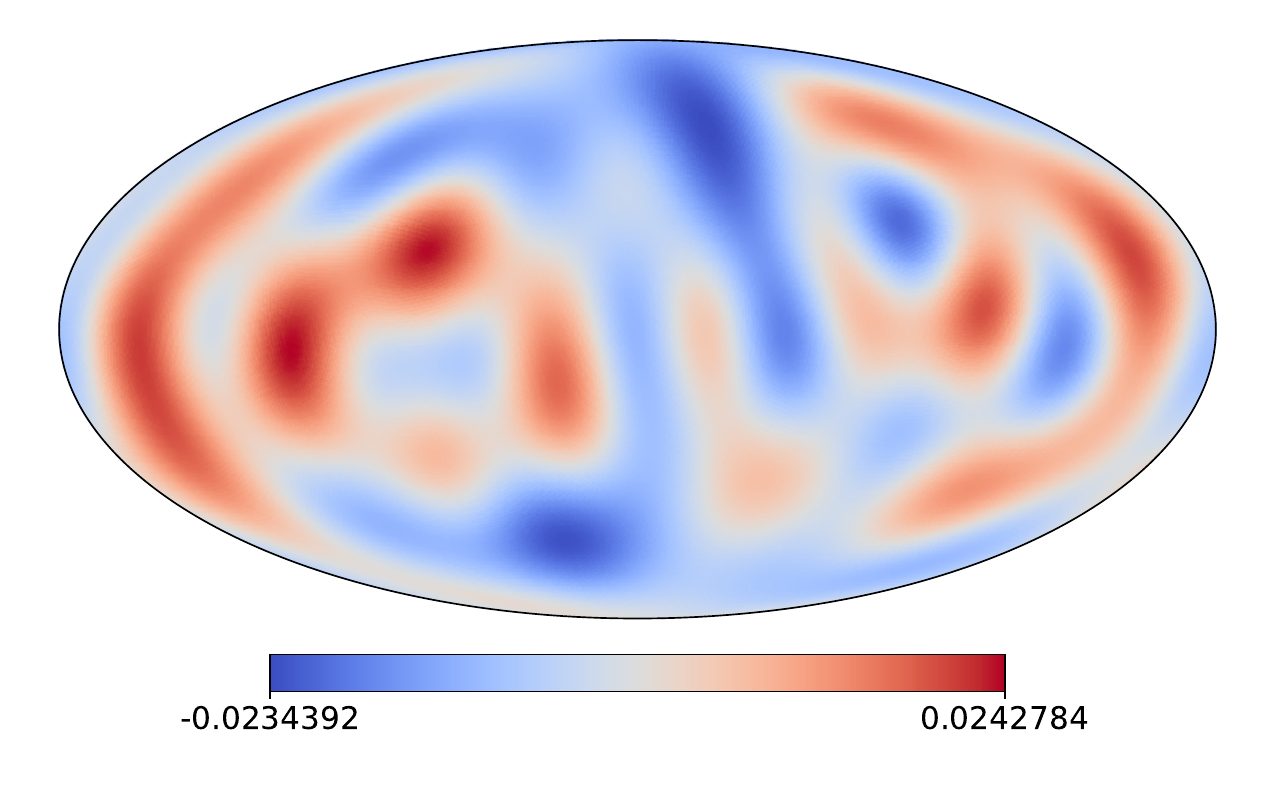}
    \includegraphics[width=0.85\linewidth]{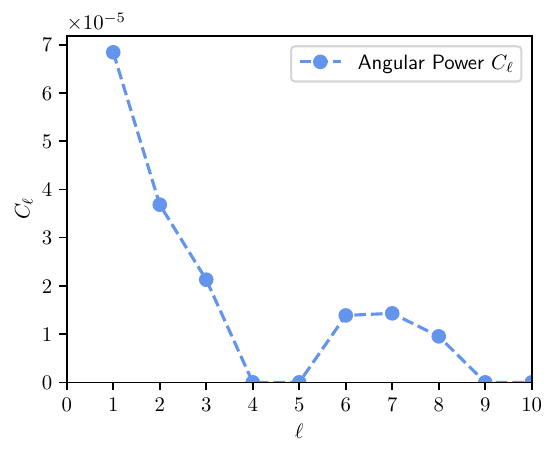}
    \caption{\textit{Top:} Multipole signal $f_{\text{mult.}}$ made by summing
            contributions from the $\ell=1$, $\ell=2$, $\ell=3$,
            $\ell=6$, $\ell=7$ and $\ell=8$ multipoles using \eqref{eq:multi_signal}.
            \textit{Bottom:} Power spectrum computed from the map above.
            \label{fig:n_multipoles}}
\end{figure}
\begin{figure}
    \centering
    \includegraphics[width=\linewidth]{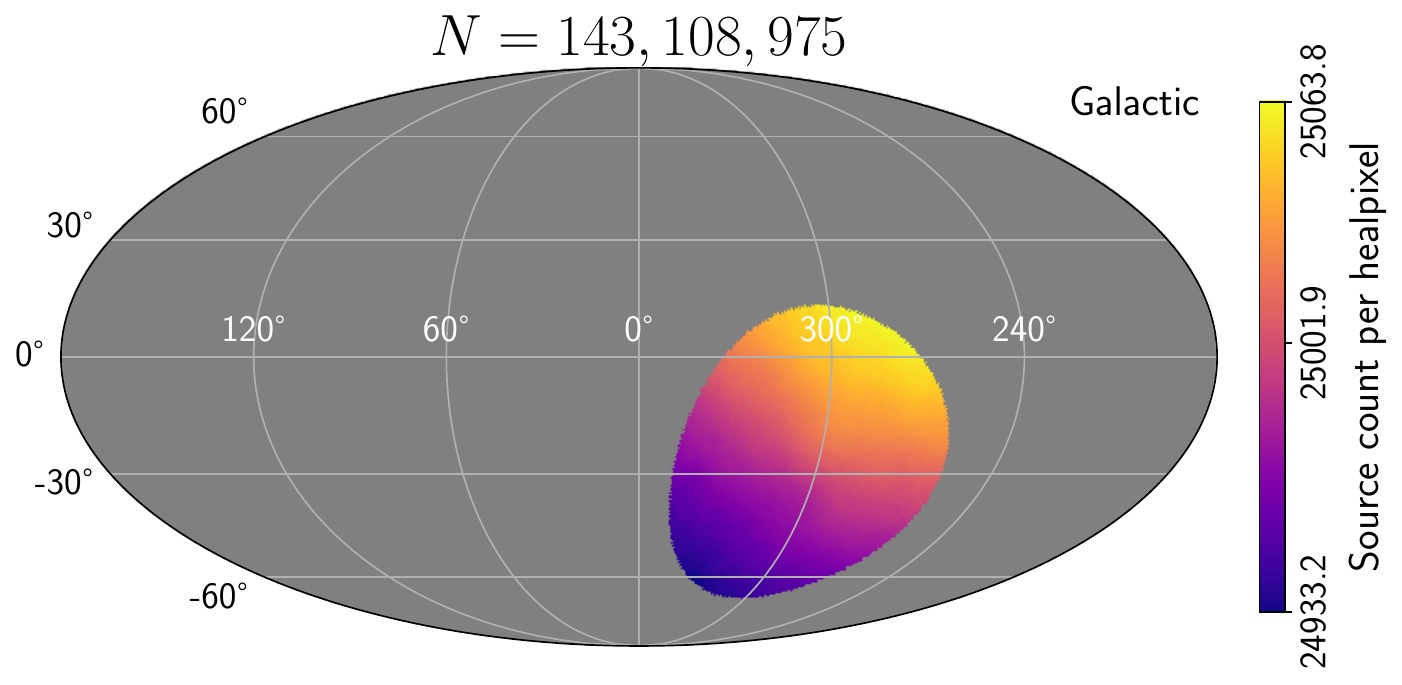}
    \includegraphics[width=\linewidth]{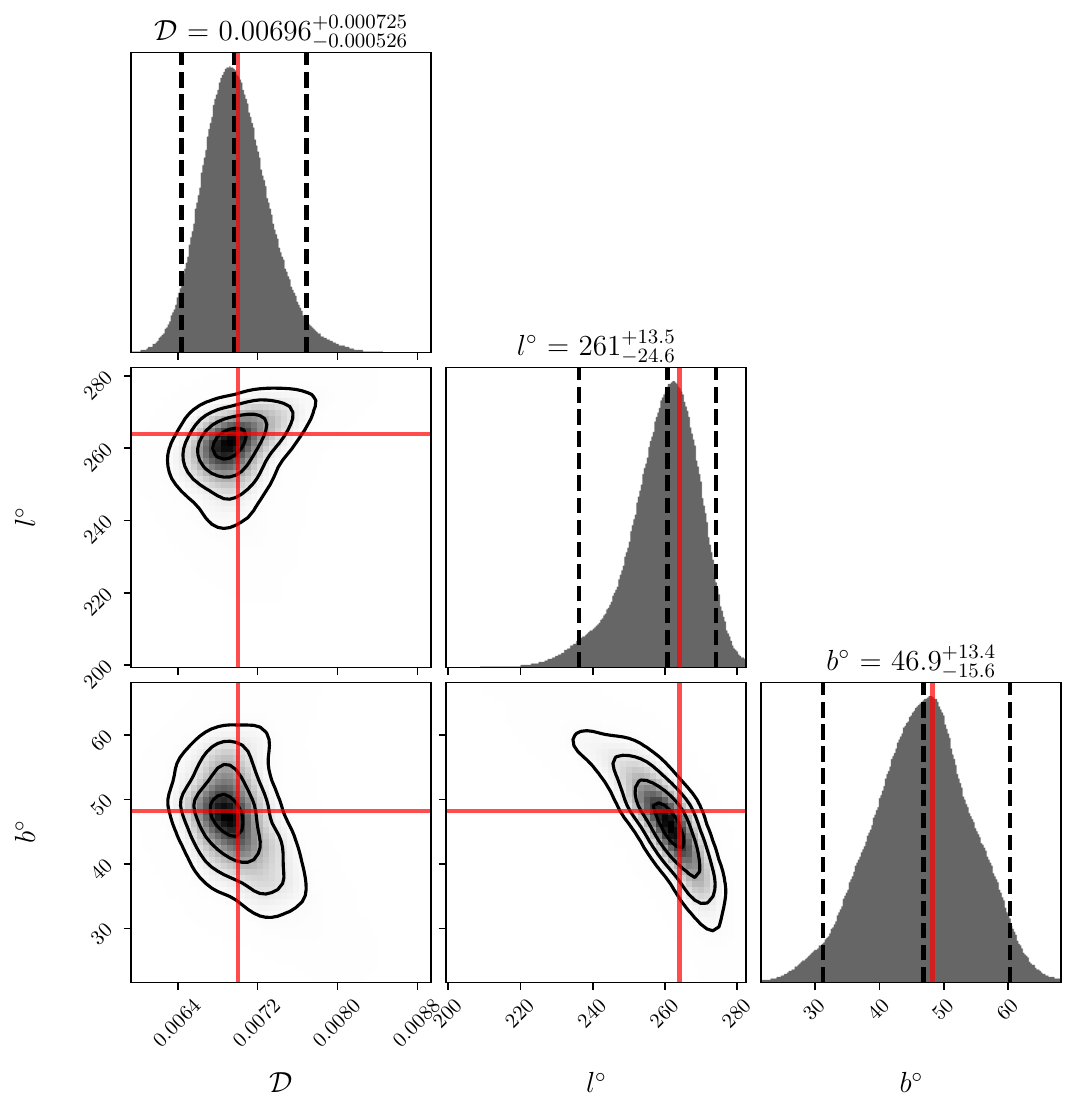}
    
    \vspace{2mm}\hfill\includegraphics[width=0.9\linewidth]{
        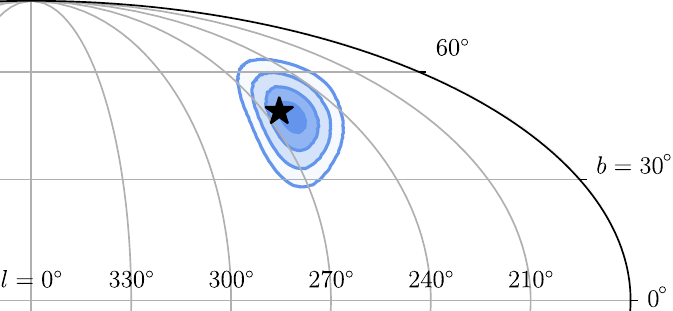}
    \caption{Results from the analysis of a small slice of sky of radius $40^\circ$
    centred at the southern equatorial pole with $N\approx 150$ million sources.
    \textit{Top:} Smoothed map, as defined in Fig.~\ref{fig:bmap_slices}.
    \textit{Middle/bottom:} Corner plot and projection of probability distribution
    for $(l^\circ, b^\circ)$ onto the sky, as defined in Fig.~\ref{fig:degen_slice}.}
    \label{fig:highN-example}
\end{figure}




\bsp	
\label{lastpage}
\end{document}